\pdfoutput=1
\documentclass[a4paper]{article}
\usepackage{amsmath}
\usepackage{amssymb}
\usepackage{slashed}
\usepackage{graphicx,epstopdf}
\usepackage{cite}
\usepackage{pbox}
\usepackage{geometry}
\usepackage{float}
\usepackage{array}
\usepackage{amsfonts}
\usepackage{graphicx}
\usepackage[font={small}]{caption}
\usepackage[plain]{fullpage}
\usepackage{xcolor}
\usepackage{hyperref}
\hypersetup{
    colorlinks=true,
    linkcolor=red,
    filecolor=magenta,      
    urlcolor=cyan,
   citecolor=blue,
} 
\definecolor{RawSienna}{cmyk}{0,0.72,1,0.45}
\definecolor{dgreen}{rgb}{0.0,0.42,0.13}
\definecolor{darkblue}{rgb}{0.0, 0.0, 0.55}
\definecolor{cornellred}{rgb}{0.7, 0.11, 0.11}
\definecolor{calpolypomonagreen}{rgb}{0.08, 0.5, 0.5}

\def\beq{\begin{equation}}
\def\eeq{\end{equation}}
\def\bea{\begin{eqnarray}}
\def\eea{\end{eqnarray}}
\makeatletter
\@addtoreset{equation}{section}

\begin{document}
\title{\LARGE \bf Baryon asymmetry via leptogenesis in a neutrino mass model with complex scaling }
\author{{Rome Samanta$^{\rm a}$\footnote{rome.samanta@saha.ac.in}, Mainak Chakraborty$^{\rm b}$\footnote{mainak.chakraborty2@gmail.com}, Probir Roy$^{\rm c}$\footnote{probirrana@gmail.com}, 
Ambar Ghosal$^{\rm a}$\footnote{ambar.ghosal@saha.ac.in}
}\\
a) Saha Institute of Nuclear Physics, HBNI, 1/AF Bidhannagar,
  Kolkata 700064, India \\  
b) Centre of Excellence in Theoretical and Mathematical Sciences,\\
SOA University, Khandagiri Square, Bhubaneswar 751030, India\\
c) Center for Astroparticle Physics and Space Science,
Bose Institute, Kolkata 700091, India  
  }
\maketitle
\begin{center}
\textbf{Abstract}
\end{center}
Baryogenesis via leptogenesis is investigated in a specific model of light neutrino masses and mixing angles. The latter was proposed on the basis of an assumed complex-extended scaling property of the neutrino Majorana mass matrix $M_\nu$, derived with a type-1 seesaw from a Dirac mass matrix $m_D$ and a heavy singlet neutrino Majorana mass matrix $M_R$. One of its important features, highlighted here, is that {\it   there is a common source of the origin of a nonzero $\theta_{13}$ and the CP violating lepton asymmetry through the imaginary part of $m_D$.}  The model predicted CP violation to be maximal for the Dirac type and vanishing for the Majorana type. We assume strongly hierarchical mass eigenvalues for $M_R$. The leptonic CP asymmetry parameter $\varepsilon^\alpha_{1}\hspace{1mm}$ with lepton flavor $\alpha$, originating from the decays of the lightest of the heavy neutrinos $N_1$ (of mass $M_1$) at a temperature $T\sim M_1$, is what matters here with the lepton asymmetries, originating from the decays of $N_{2,3}$, being washed out. The light leptonic and heavy neutrino number densities (normalized to the entropy density) are evolved via Boltzmann equations down to electroweak temperatures to yield a baryon asymmetry through sphaleronic transitions. The effects of flavored vs. unflavored leptogenesis in the three mass regimes (1) $M_1<10^{9}$ GeV, (2) $10^9$ GeV $<M_1<$ $10^{12}$ GeV and (3) $M_1>10^{12}$ GeV are numerically worked out for both  a normal and  an inverted mass ordering of the light neutrinos. Corresponding results on the baryon asymmetry of the universe are obtained, displayed and discussed. For  values close to the best-fit points of the input neutrino mass and mixing parameters, obtained from neutrino oscillation experiments, successful baryogenesis is achieved for the mass regime (2) and a normal mass ordering of the light neutrinos with a nonzero $\theta_{13}$ playing a crucial role. However, the other possibility of an inverted mass ordering for the same mass regime, though disfavored, cannot be excluded. A discussion is also given on the sensitivity of our result to the masses $M_{2,3}$ of the heavier neutrinos $N_{2,3}$.
\newpage
\section{Introduction}\label{s1}
Much effort has already been made towards understanding the origin of the baryon asymmetry of the universe $Y_B=(n_B-n_{\bar{B}})/s\simeq(8.7\pm0.1)\times10^{-11}$\cite{1} -- the number density ($n_B$) of baryons minus that ($n_{\bar{B}}$) of antibaryons normalized to the  entropy density $s$. A comprehensive review with references may be found in Ref. \cite{2}. Various possible mechanisms have been considered for this purpose, e.g. GUT baryogenesis, electroweak baryogenesis, the Affleck-Dine mechanism and baryogenesis via leptogenesis. We concentrate on the last-mentioned possibility \cite{3,4,5,6,7,8}. Here a CP odd particle-antiparticle asymmetry is first generated at a high scale in the leptonic sector; that is thereafter converted into a baryon asymmetry  by sphaleron processes during the electroweak phase transition. In the most popular extension of the Standard Model (SM) for generating light neutrino masses, three\footnote{This can be done with two heavy RH singlet neutrinos but not with just one.} heavy right-chiral (RH) singlet neutrinos are added to induce tiny neutrino masses and their mixing angles through the type-1 seesaw mechanism \cite{9,10,11,12}. The complex Yukawa couplings $f_{i\alpha}^N$, that connect those singlet RH neutrinos $N_i$ to the SM-doublet left-chiral leptons of flavor $\alpha$, generate the necessary CP violation in the decays of those heavy RH neutrinos into the Higgs scalar plus the SM leptons. The occurrence of  Majorana mass terms for the heavy neutrinos in the Lagrangian provides the required lepton nonconservation. The rate of interaction with those Yukawa couplings being smaller than the Hubble expansion rate, departure from thermal equilibrium  ensues. Hence all the Sakarav conditions \cite{13} are fulfilled for generating $Y_B$. The present work is devoted to a quantitative study of the origin of $Y_B$ via leptogenesis in a model \cite{14,add} of neutrino masses with complex scaling -- proposed by some of us. As a step towards that, we shall summarize  the relevant features of the concerned model in the next Sec. \ref{s2}.

\paragraph{}
First, let us establish our notation and convention by choosing without loss of generality the Weak Basis (sometimes called the leptogenesis basis \cite{15}) in which the $3\times3$ mass matrices, not only  of the charged leptons but also of the heavy RH neutrinos, are diagonal with nondegenerate real and positive entries, e.g. $M_R={\rm diag}\hspace{1mm}(M_1,M_2,M_3)$, $M_i\hspace{1mm}(i=1,2,3)>0$. We shall work in the strongly hierarchical  scenario in the right-chiral neutrino sector in which those masses will be taken to be widely spaced. Specifically, we assume that $M_1<<M_2<<M_3$. A crucial input into these scenarios is the flavor structure of the neutrino Dirac mass matrix $m_D$. The latter appears in the neutrino mass terms of the Lagrangian as  
\bea
-\mathcal{L}_{mass}^{\nu,N}=\bar{N}_{iR} (m_D)_{i\alpha}\nu_{L\alpha}+\frac{1}{2}\bar{N}_{iR}(M_R)_i \delta _{ij}N_{jR}^C + {\rm h.c}. \label{selag}
\eea
with $N^C_j=C\bar{N}_j^T$. The effective light neutrino Majorana mass matrix $M_\nu$ is then given by the standard seesaw result \cite{9,10,11,12}
\bea
M_\nu = -m_D^TM_R^{-1}m_D. \label{swmnu}
\eea
This $M_\nu$ enters the effective low energy neutrino mass term in the Lagrangian as
\bea
-\mathcal{L}_{mass}^\nu= \frac{1}{2}\bar{\nu_{L\alpha}^C} (M_\nu)_{\alpha\beta}\nu_{L\beta} + {\rm h.c}. \label{lag}
\eea 
It is a complex symmetric $3\times3$ matrix ($M_\nu^*\neq M_{\nu}=M_\nu^T$) which can be put into a diagonal form by a similarity transformation with a unitary matrix $U$:
\bea
U^T M_\nu U=M_\nu^d \equiv \rm diag\hspace{1mm}(m_1,m_2,m_3)\label{e0}
\eea  
with ${\rm m}_i\hspace{1mm}(i=1,2,3)$ taken to be nonzero, real and small positive masses $<$ $\mathcal{O}({\rm eV})$. In our Weak Basis we can take $U$ as 
\bea
U=U_{PMNS}\equiv 
\begin{pmatrix}
c_{1 2}c_{1 3} & e^{i\frac{\alpha}{2}} s_{1 2}c_{1 3} & s_{1 3}e^{-i(\delta - \frac{\beta}{2})}\\
-s_{1 2}c_{2 3}-c_{1 2}s_{2 3}s_{1 3} e^{i\delta }& e^{i\frac{\alpha}{2}} (c_{1 2}c_{2 3}-s_{1 2}s_{1 3} s_{2 3} e^{i\delta}) & c_{1 3}s_{2 3}e^{i\frac{\beta}{2}} \\
s_{1 2}s_{2 3}-c_{1 2}s_{1 3}c_{2 3}e^{i\delta} & e^{i\frac{\alpha}{2}} (-c_{1 2}s_{2 3}-s_{1 2}s_{1 3}c_{2 3}e^{i\delta}) & c_{1 3}c_{2 3}e^{i\frac{\beta}{2}} 
\end{pmatrix}\label{eu}
\eea 
with $c_{ij}\equiv\cos\theta_{ij}$, $s_{ij}\equiv\sin\theta_{ij}$ and $\theta_{ij}=[0,\pi/2]$. CP violation enters here through nonzero values of the Dirac phase $\delta$ and of the Majorana phases $\alpha,\beta$  with $\delta,\alpha,\beta=[0,2\pi]$. We follow the PDG convention \cite{16} on these angles and phases except that we denote the Majorana phases by $\alpha$ and $\beta$.
 
\paragraph{}
In the main body of the paper we calculate  the CP asymmetry originating from the decays $N_i\rightarrow \slashed{L}_\alpha \phi$, $\slashed{L}_\alpha^C \phi^\dagger$ where $\slashed{L}_\alpha$ and $\phi$  are the respective fields of the SM left-chiral lepton doublet of flavor $\alpha$ and the Higgs doublet. This is done in terms of the imaginary parts of  appropriately  defined quartic products of the neutrino Dirac mass matrix $m_D$ and its hermitian conjugate $m_D^{\dagger}$, as well as of an explicit function of the variable $x_{ij}\equiv M_j^2/M_i^2$. Clearly, the calculation depends sensitively  on the flavor structure of $m_D$ and hence on the specific neutrino mass model under consideration. The CP asymmetries (and therefore leptogenesis as a whole) may be flavor dependent or independent according to the temperature regime in which the CP violating decays take place. For an evolution down to the electroweak scale, one needs to solve the corresponding Boltzmann Equations.  We therefore consider the Boltzmann evolution equation for the number density $n_a$ of a particle of type $a$ (either a right-chiral heavy neutrino $N_i$ or a left-chiral lepton doublet $\slashed{L}_\alpha$)  normalized to the photon number density $n_\gamma$. For this purpose, we take
\bea
\eta_a(z)=\frac{n_a(z)}{n_\gamma(z)},
 n_\gamma(z)=\frac{2 M_{1}^3}{\pi^2 z^3} 
\eea \label{z}
as functions of $z\equiv$ $M_1/T$. We rewrite these equations for the variable $Y_a$ where
\bea
Y_a = n_a/s=\frac{n_\gamma}{s}\eta_a=1.8g_{*s} \eta_a,
\eea 
$g_{*s}$ being the total number of effective and independent massless degrees of freedom at the concerned temperature. The evolution of $Y_a$ is studied for different $a$'s from a temperature of the order of the lightest right-chiral neutrino mass $M_{1}$ to that of the electroweak phase transition where sphaleron-induced processes take place converting the lepton asymmetry into  a baryon asymmetry $Y_B$.
\paragraph{}
In pursuing $Y_B$, we need to zero in on $Y_{\Delta_{\lambda}}$ where $\Delta_\lambda=\frac{1}{3}B-L_\lambda$ with $B$ being the baryon number and $L_{\lambda}$ the lepton number of the active flavor $\lambda$. The analysis is done numerically but in three different regimes \cite{7, 8} depending on where $M_1$ lies: (1) $M_{1}<10^9$ GeV where all the lepton flavor are distinctly active, (2)  $10^9\hspace{1mm}{\rm GeV}<M_{1}<10^{12}\hspace{1mm}{\rm GeV}$ where $e$ and $\mu$ flavors are indistinguishable but the $\tau$-flavor is separately active and (3) $M_{1}>10^{12}~{\rm GeV}$ where all lepton flavors are indistinguishable. The quantity $Y_B$ and $Y_{\Delta_{\lambda}}$ are linearly related but with different numerical coefficients for the three different regimes. In our  numerical analysis, six constraints from experimental and observational data are inputted: the 3$\sigma$ ranges of the solar and atmospheric neutrino mass squared differences as well as of the three neutrino mixing angles plus the cosmological upper bound on the sum of the three light neutrino masses.  The analysis is done separately in each regime for a normal mass ordering ($m_3>m_2>m_1$) as well as for an inverted ordering ($m_2>m_1>m_3$) of the light neutrinos. The final results are tabulated numerically  as well as displayed  graphically.\\

We have already mentioned the content of Section \ref{s2}. The rest of the paper is organized as follows. In Section \ref{s3} we calculate the CP asymmetry parameters generated in the decays of $N_i$ into $\slashed{L}_\alpha \phi$ and $\slashed{L}_\alpha^C \phi^\dagger$. Section \ref{s4} contains an algebraic treatment of the Boltzmann evolution equations and  of the generation of the baryon asymmetry $Y_B$ in the three mass regimes. The numerical analysis that follows is detailed with a discussion of its consequences in Section \ref{s5}. Section \ref{s6} addresses the possible role played by the heavier neutrinos $N_{2,3}$. A summary of our work is given in the last Section \ref{s7}.
\section{Complex scaling with type-I seesaw}\label{s2}

A key feature of $M_\nu$ is the $\mathbb{Z}_2\times \mathbb{Z}_2$ residual symmetry \cite{17} that it possesses.  This is an invariance of $M_\nu$ under a linear transformation on neutrino fields
\bea
\nu_{L\alpha}\rightarrow G_{\alpha\beta}\nu_{L\beta},
\eea
i.e. in  a matrix notation, with a $3\times 3 $ matrix $G$,
\bea
G^T M_\nu G=M_\nu.
\eea
One can show \cite{17} that there are two independent matrices $G^{2,3}$ implementing this invariance and obeying the unitary diagonalization 
\bea
G^{2,3}U=Ud^{2,3},
\eea
where $d^2={\rm diag.}\hspace{1mm} (-1,1-,1)$ and  $d^3={\rm diag.}\hspace{1mm}(-1,-1,1)$. Some of us have proposed \cite{14} a complex extension of this symmetry by considering the nonstandard CP transformations  
\bea
\nu_{L\alpha }\rightarrow i (G_L)_{\alpha \beta}\gamma^0 \nu_{L\beta}^C, \hspace{.5cm}
N_{Ri }\rightarrow i (G_R)_{ij}\gamma^0 N_{Rj}^C \label{seCP}
\eea
and demanding the invariance relations 
\bea
G_R^{\dagger}m_D G_L=m_D^*, \hspace{.3cm} G_R^{\dagger}M_R G_R^*=M_R^*. \label{trmdmr}
\eea

Eqs. (\ref{swmnu}) and (\ref{trmdmr}) together imply \bea G_L^TM_\nu G_L=M_\nu^* \eea
which is our complex-extended invariance statement on the low energy neutrino Majorana mass matrix $M_\nu$. At this point, $G_L$ is taken to be \cite{14}
\bea
G_L=G_3^{{\rm scaling}}=
\begin{pmatrix}
-1&0&0\\0&(1-k^2)(1+k^2)^{-1}&2k(1+k^2)^{-1}\\
0&2k(1+k^2)^{-1}&-(1-k^2)(1+k^2)^{-1}
\end{pmatrix}=(G_3^{scaling})^T,\label{e15}
\eea 
$k$ being a real scaling factor. This $G_3^{{\rm scaling}}$ is the operative residual symmetry generator for the original scaling ansatz \cite{18,19,a1,a2}. It now obeys the relation
\bea
G_3^{{\rm scaling}}U^*=U\tilde{d},
\eea
where $\tilde{d}_{\alpha\beta}$ equals $\pm \delta_{\alpha\beta}$ and hence admits eight possibilities.  Only four of these were shown\cite{14} to be viable and led independently to the results
\bea
\tan\theta_{23}=k^{-1}, \label{atm}
\eea
\bea
\sin\alpha=\sin\beta=\cos\delta=0.\label{Cpv}
\eea
The detailed phenomenological consequences of (\ref{atm}) and (\ref{Cpv}) were worked out in Ref. \cite{14}. The most general $M_\nu$, that satisfies
\bea  
 (G_3^{{\rm scaling}})^T M_\nu G_3^{{\rm scaling}}=M_\nu^*, \label{ces}
 \eea
is given by the complex-extended scaling (CES) form of $M_\nu$, namely \cite{14}
\bea
 M_\nu^{CES} = \begin{pmatrix}
 x& -y_1 k +iy_2k^{-1}&y_1+iy_2\\
 -y_1 k +iy_2k^{-1}&z_1-wk^{-1}(k^2-1)-iz_2&w-iz_2(2k)^{-1}(k^2-1)\\
 y_1+iy_2 &w-iz_2(2k)^{-1}(k^2-1)&z_1+i z_2
\end{pmatrix}, \label{e3}
\eea
where $x$, $y_{1,2}$, $z_{1,2}$ and $ w$ are real mass dimensional quantities. 

Since  $M_R$ has been taken  to be diagonal, the corresponding symmetry generator matrix $G_R$, cf. the second of Eqs. (\ref{trmdmr}), is diagonal with entries $\pm1$, i.e.
\bea
G_R=\rm diag \hspace{1mm}(\pm1,\pm1,\pm1).\label{G_R}
\eea
Thus there are eight different structures of $G_R$. Correspondingly, from the first relation of (\ref{trmdmr}), there could be eight possible different structures of $m_D$. It can be shown by tedious algebra that all other structures of $G_R$, except for \bea G_R=\rm diag \hspace{1mm} (-1,-1,-1), \label{gr}\eea  are incompatible with  scaling symmetry \cite{18}. Thus we take  $G_R$ of (\ref{gr}) as the only viable residual symmetry of $M_R$. We can now write the first of (\ref{trmdmr}) as
\bea
m_DG_L=-m_D^* \label{5p7}
\eea
which is really a complex extension of the Joshipura-Rodejohann result\footnote{Those authors followed a different phase convention; they obtained $m_DG_L=m_D$ instead of $m_DG_L=-m_D$.}\cite{20}
$m_DG_L=-m_D.$\\

 The most general form of $m_D$ that satisfies (\ref{5p7}) is 
 \bea
 m_D^{CES}= \begin{pmatrix}
 a & b_1+ib_2 &-b_1/k+ib_2k\\
 e & c_1+ic_2 & -c_1/k+ic_2k\\
 f & d_1+id_2 & -d_1/k+id_2k\\
 \end{pmatrix}, \label{mdces}
 \eea
 where $a$, $b_{1,2}$, $c_{1,2}$, $d_{1,2}$, $e$ and $f$ are nine a priori unknown  real mass dimensional quantities apart from the real, positive, dimensionless $k$. Using (\ref{swmnu}), $M_\nu^{CES}$ of (\ref{e3})  obtains  with the real mass parameters  $x$, $y_{1,2}$, $z_{1,2}$ and $ w$ related to those of (\ref{mdces}), as given in Table \ref{t}. It is noteworthy that whereas $m_D^{CES}$ has ten real parameters, $M_\nu^{CES}$ has only seven.
\begin{table}[H]
\begin{center}
\caption{Parameters of $M_\nu^{CES}$ in terms of the parameters of $m_D$ and $M_R$.} \label{t}
 \begin{tabular}{|c|} 
\hline 
$x=-(\frac{a^2}{M_1}+\frac{e^2}{M_2}+\frac{f^2}{M_3})$\\
$y_1=\frac{1}{k}(\frac{ab_1}{M_1}+\frac{ec_1}{M_2}+\frac{fd_1}{M_3})$\\
$y_2=k(\frac{ab_2}{M_1}+\frac{ec_2}{M_2}+\frac{fd_2}{M_3})$\\
$z_1=-\frac{1}{k^2}(\frac{b_1^2}{M_1}+\frac{c_1^2}{M_2}+\frac{d_1^2}{M_3})+k^2(\frac{b_2^2}{M_1}+\frac{c_2^2}{M_2}+\frac{d_2^2}{M_3})$\\
$z_2=\frac{2b_1b_2}{M_1}+\frac{2c_1c_2}{M_2}+\frac{2d_1d_2}{M_3}$\\
$w=\frac{1}{k}(\frac{b_1^2}{M_1}+\frac{c_1^2}{M_2}+\frac{d_1^2}{M_3})+k(\frac{b_2^2}{M_1}+\frac{c_2^2}{M_2}+\frac{d_2^2}{M_3})$\\
\hline
\end{tabular} 
\end{center} 
\end{table}
 One can count the real parameters, as given in $m_D$ of (\ref{mdces}). Along with the RH neutrino masses $M_1$, $M_2$, $M_3$, one obtains a set of thirteen real parameters for $M_\nu$. In order to reduce the number of parameters towards attaining the goal of a tractable result, we first use the assumed hierarchical nature of the RH neutrino masses $M_1<<M_2<<M_3$. We then take the parameters $d_{1,2}$, $e$ and $f$ in Table \ref{t} to be of the same order of magnitude as $a$, $b_{1,2}$ and $c_{1,2}$. That enables us to neglect all terms in Table \ref{t} with $M_3$ in the denominator. Now we rescale the remaining parameters of Table \ref{t} as follows:
 \bea
 a\longrightarrow a^\prime=\frac{a}{\sqrt{M_1}},\label{rsc1}\\
  b_{1,2}\longrightarrow b_{1,2}^\prime=\frac{b_{1,2}}{\sqrt{M_1}},\\
   c_{1,2}\longrightarrow c_{1,2}^\prime=\frac{c_{1,2}}{\sqrt{M_2}},\\
    e\longrightarrow e^\prime=\frac{e}{\sqrt{M_2}}.\label{rsc2}
 \eea
 Consequently, the entries of Table \ref{t} can be written in terms of the rescaled parameters as in Table \ref{tres}. We are now left with a six-dimensional parameter space with the real parameters $x$, $y_{1,2}$, $z_{1,2}$ and $w$ as given in Table \ref{tres}. Note that, had we neglected the terms with $M_2$ in the denominator too, we would have been left with a three dimensional parameter space which would have been in a danger of being overdetermined by the six experimental and observational constraints mentioned in the Introduction. We shall latter discuss how to estimate the missing parameters $f$ and $d_{1,2}$. 
 
 \begin{table}[H]
\begin{center}
\caption{Parameters of $m_D^{CES}$ in the rescaled version.} \label{tres}
 \begin{tabular}{|c|} 
\hline
$x=-({a^\prime}^2+{e^\prime}^2)$ \\
 $y_1=\frac{1}{k}(a^\prime b_1^\prime +e^\prime c_1^\prime)$\\
 $y_2=-k(a^\prime b_2^\prime+ e^\prime c_2^\prime) $\\
$ z_1=-\frac{1}{k^2}({b_1^\prime}^2 +{c_1^\prime}^2) +k^2({b_2^\prime}^2 +{c_2^\prime}^2)$\\
 $z_2= 2 b_1^\prime b_2^\prime +2 c_1^\prime c_2^\prime $\\
 $w= \frac{1}{k}({b_1^\prime}^2 +{c_1^\prime}^2) +k({b_2^\prime}^2 +{c_2^\prime}^2)$\\
\hline
\end{tabular} 
\end{center} 
\end{table}

Before concluding this section, let us make an important point. In the absence of of any imaginary part of the matrix $m_D^{CES}$ of  (\ref{mdces}), the seesaw relation (\ref{swmnu}) gives rise to the Generalized Real Scaling form of $M_\nu$, namely \cite{14}
\bea
M_\nu^{GRS}=\begin{pmatrix}
x&-y_1k&y_1\\-y_1k&z_1-wk^{-1}(k^2-1)&w\\y_1&w&z_1
\end{pmatrix}\label{e19}
\eea 
 with real mass-dimensional entries. However, as was explained in Ref. \cite{14}, in this case $\theta_{13}$ vanishes and so information about the Dirac CP violating phase $\delta$ is lost. Moreover, owing to the real nature of the associated $m_D^{GRS}$, there is no Majorana CP violation either. Thus we see that the imaginary part of $m_D^{CES}$ is the common source of an operative nonzero $\theta_{13}$ as well as CP violation in leptonic sector. The latter is in fact crucial to leptogenesis which is effected through a nonzero value of the CP asymmetry parameter $\epsilon$, as explained in the next section. It is through the nonvanishing nature of Im$\hspace{1mm} m_D^{CES}$ that the final matter-antimatter asymmetry in the universe gets directly related to the low energy parameters  $\theta_{13}$ and $\delta$. 
\section{Calculation of CP asymmetry parameter} \label{s3}

The part of our  Lagrangian relevant to the generation of a CP asymmetry is
\bea
  -\mathcal{L}_{D} = f^{N}_{i \alpha} \overline{N}_{Ri}\tilde{\phi}^\dagger \slashed{L}_\alpha  + h.c. , \label{decay}
\eea
where $\slashed{L}_\alpha =(\nu_{L_\alpha}~\ell^{-}_{L\alpha})^T$ is the left-chiral SM lepton doublet of flavor $\alpha$, while   $\tilde{\phi}=(\phi^{0*}~-\phi^{-})^T$ is the charge conjugated Higgs scaler
doublet. It is evident from  (\ref{decay}) that the decay products of $N_i$ can be $\ell_\alpha^-\phi^+,\nu_\alpha\phi^0,\ell_\alpha^+\phi^-$ and $\nu_\alpha^C\phi^{0*}$. We are interested in the flavor dependent CP asymmetry parameter $\varepsilon^\alpha_i $ which is given by 
\begin{eqnarray}
 \varepsilon^\alpha_i &=&\frac{\Gamma({N}_i\rightarrow
    \slashed{L}_\alpha \phi)-\Gamma({N}_i\rightarrow
   \slashed{L}^C_\alpha \phi^{\dagger})}{\Gamma({N}_i\rightarrow
    \slashed{L}_\alpha \phi)+\Gamma({N}_i\rightarrow
   \slashed{L}^C_\alpha \phi^{\dagger})},\label{cpa}
\end{eqnarray}
 $\Gamma$ being  the corresponding  partial decay width.  A nonzero value of $\varepsilon^\alpha_i$ needs to  arise out of the interference between the tree level and one loop contributions\cite{4}. This is since at the tree level we have 
 \bea
 \Gamma^{tree}(N_i\rightarrow \slashed{L}_\alpha \phi)=\Gamma^{tree}(N_i\rightarrow \slashed{L}^C_\alpha \phi^{\dagger})=(16\pi)^{-1}(f_{i\alpha}^{N\dagger}f_{i\alpha}^{N})M_i,\hspace{1mm} {\rm (no\hspace{1mm}sum\hspace{1mm}over\hspace{1mm}i}).\eea 
 
\begin{figure}
\begin{center}
\includegraphics[scale=0.5]{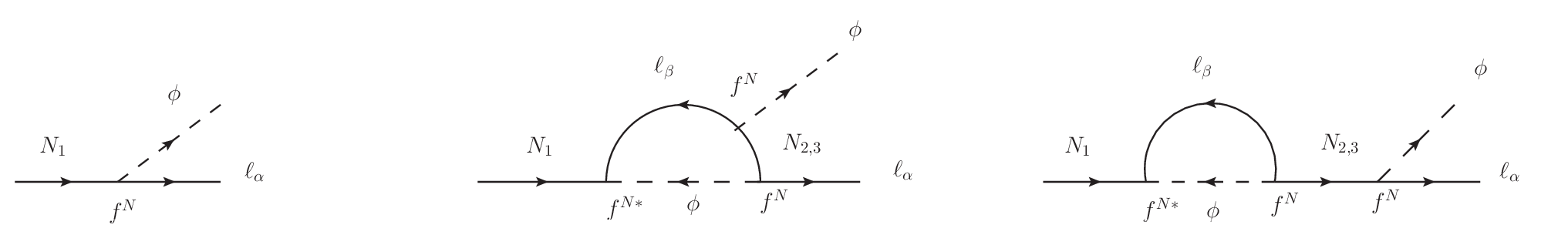}
\caption{Tree level as well as one loop vertex correction and self energy diagrams that contribute to the CP asymmetry parameter $\varepsilon_1^{\alpha}$. The flavor of the internal charged lepton $\ell_{\beta}$ is summed and the Yukawa coupling $f^N$ is supplied with appropriate flavor indices in the interference amplitude.}\label{cpdiag}
\end{center}
\end{figure}

 One loop contributions come both from vertex correction and self-energy terms (cf. Fig.\ref{cpdiag}). For leptogenesis with hierarchical heavy RH neutrinos, (\ref{cpa})  can be evaluated to be
\bea
 \varepsilon_i^{\alpha}=\frac{1}{4\pi v^2 \mathcal{H}_{ii}}\sum\limits_{j \neq i} g(x_{ij})\hspace{1mm} {\rm Im}\hspace{1mm}\mathcal{H}_{ij}(m_D)_{i \alpha}(m_D^*)_{j \alpha}+\frac{1}{4\pi v^2 {\mathcal{H}}_{ii}}\sum_{j\ne i}\frac{ {\rm Im}\hspace{1mm}{\mathcal{H}}_{ji}(m_D)_{i \alpha }({m_D}^*)_{ j \alpha }}{(1-x_{ij})}.
\label{epsi_intro_h}
\eea
In (\ref{epsi_intro_h}) $<\phi^0>=v/\sqrt{2}$ so that $m_D=vf^N/\sqrt{2}$, $\mathcal{H}\equiv m_D {m_D}^\dagger$ and
$x_{ij}$ was defined in Sec.\ref{s1}. Furthermore, $g(x_{ij})$ is given by
\bea
 g(x_{ij})&=&\frac{\sqrt{x_{ij}}}{1-x_{ij}}+f(x_{ij}),\label{loop}
 \eea
where the first RHS term arises from the one loop self energy term interfering with the tree level contribution. The second RHS term in (\ref{loop}), originating from the interference of the contribution from the one loop vertex correction diagram with the tree level term, is given by
\bea
f(x_{ij})=\sqrt{x_{ij}}\left[ 1-(1+x_{ij}){\rm ln}\left(\frac{1+x_{ij}}{x_{ij}}\right) \right].\label{loop2}
\eea

Let us discuss some physics aspects of (\ref{epsi_intro_h}). As already mentioned, depending upon the temperature regime in which leptogenesis occurs, lepton flavors may be fully distinguishable, partly distinguishable or indistinguishable. It is reasonable to assume that leptogenesis takes place at $T\sim M_{1}$. It is known \cite{4} that  lepton flavors cannot be treated separately if the concerned  process occurs above a temperature $T\sim M_{1}> 10^{12}~{\rm GeV}$. In case the said temperature is lower, two possibilities arise.  When $T\sim M_{1}< 10^9$ GeV all three ($e,\mu,\tau$) flavors are individually
active and we need three CP asymmetry parameters $\varepsilon^e_i,\varepsilon^\mu_i,\varepsilon^\tau_i$
for each generation of RH neutrinos. On the other hand when we have $10^9~{\rm GeV}<T\sim M_{1}< 10^{12}~{\rm GeV}$, only the $\tau$-flavor can be identified separately while the $e$ and $\mu$ act indistinguishably. Here we need two CP asymmetry parameters $\varepsilon^{(2)}_i=\varepsilon^e_i+\varepsilon^\mu_i$ and
$\varepsilon^\tau_i$ for each  of the RH neutrinos. As an aside, let us point out a simplification in this model for  unflavored leptogenesis which is relevant for the high temperature regime. Summing over all $\alpha$,

\begin{equation}
\sum\limits_{\alpha} {\rm Im} \hspace{1mm}\mathcal{H}_{ji}(m_D)_{i \alpha }({m_D}^*)_{j \alpha }
={\rm Im} \hspace{1mm}{H}_{ji} {\mathcal{H}}_{ij}={\rm Im} \hspace{1mm}{H}_{ji} {{\mathcal{H}}^*_{ji}}={\rm Im} \hspace{1mm}|{\mathcal{H}}_{ji}|^2=0,
\end{equation}
i.e. the second term in the RHS of (\ref{epsi_intro_h}) vanishes. The flavor-summed CP asymmetry parameter is therefore given by the simplified expression
\begin{eqnarray}
\varepsilon_i &=& \sum\limits_{\alpha}\varepsilon^\alpha_i \nonumber\\
              &=& 
\frac{1}{4\pi v^2 \mathcal{H}_{ii}}\sum_{j\ne i} 
\left[f(x_{ij})+\frac{\sqrt{x_{ij}}}
{(1-x_{ij})}\right] {\rm Im}\hspace{1mm}{\mathcal{H}_{ij}}{\mathcal{H}_{ij}}.
\label{sum_epsi_intro_h2}
\end{eqnarray}
In the mass model \cite{14} being considered, it follows from (\ref{mdces}) that 
\bea
\mathcal{H}^{CES}=\begin{pmatrix}
a^2+b_1^2p+b_2^2q & ac+b_1c_1p+b_2c_2q & af+b_1d_1p+b_2d_2q\\
ac+b_1c_1p+b_2c_2q & e^2+c_1^2p+c_2^2q & ef+c_1d_1p+c_2d_2q\\
af+b_1d_1p+b_2d_2q & ef+c_1d_1p+c_2d_2q & f^2+d_1^2p+d_2^2q
\end{pmatrix}\label{hces}
\eea
with $p=1+k^{-2}$ and $q=1+k^{2}$. Since (\ref{hces}) implies that  Im $\mathcal{H}^{CES}$=0, it follows from (\ref{sum_epsi_intro_h2}) that 
\bea
\varepsilon_i=0,
\eea
i.e. flavored-summed leptogenesis does not take place for any $N_i$. With the assumption that only the decay of $N_1$ matters in generating the CP asymmetry, $\varepsilon_1$ is the pertinent quantity for unflavored leptogenesis, but it vanishes. This nonoccurrence of unflavored leptogenesis is one of the robust predictions of the model.  
\paragraph{} 
 Next, we focus on the calculation of the $\alpha$-flavored CP asymmetry in terms of $x_{12}$, $x_{13}$ and the elements of $m_D^{CES}$. These  are relevant for the fully flavored as well as the
$\tau$-flavored regimes. We find that
\bea
&& \varepsilon^e_1=0, 
\eea
while
\bea
&& \varepsilon^\mu_1= \xi[b_2k^2(\chi_1+\chi_2)+b_1(\chi_3+\chi_4)-b_1^2\chi_5]=-\varepsilon^\tau_1. \label{epslnsn}
\eea
In (\ref{epslnsn}) the real parameters $\xi$ and $\chi_{i}\hspace{1mm}{(i=1-5)}$ are defined as
\bea
\xi &=&\frac{1}{4[b_1^2+(a^2+b_1^2+b_2^2)k^2+b_2^2k^4]\pi v^2},\nonumber\\
\chi_1 &=& b_2(1+k^2)[c_1c_2\{1+g(x_{12})-x_{12}\}+d_1d_2\{1+g(x_{13})-x_{13}\}],\nonumber\\
\chi_2 &=& a[c_1e\{1+g(x_{12})-x_{12}\}+d_1f\{1+g(x_{13})-x_{13}\}],\nonumber\\
\chi_3 &=& b_2(1+k^2) [ c_1^2\{1+g(x_{12})-x_{12}\}-k^2 [c_2^2\{1+g(x_{12})\nonumber\\&-&x_{12}\}+d_2^2\{1+g(x_{13})-x_{13}\} ]+d_1^2\{1+g(x_{13})-x_{13}\}],\nonumber\\
\chi_4 &=&-ak^2[c_2e\{1+g(x_{12})-x_{12}\}+d_2f\{1+g(x_{13})-x_{13}\}],\nonumber\\
\chi_5 &=&  k^2[c_1c_2\{1+g(x_{12})-x_{12}\}+d_1d_2\{1+g(x_{13})-x_{13}\}]. \label{xi}
\eea
Thus the nonzero leptonic CP asymmetry parameter  $\varepsilon_1^\mu=-\varepsilon_1^\tau$ depends on all ten parameters of $m_D^{CES}$ as well as on $x_{12}$ and $x_{13}$.
\paragraph{}
We had earlier identified Im $m_D^{CES}$ as the common source of the origin of a nonzero $\theta_{13}$ and leptonic CP violation. A real $m_D^{CES}$ implies vanishing values for $b_2$, $c_2$ and $d_2$ in which case $\varepsilon_1^\mu=-\varepsilon_1^\tau$ vanishes identically and, as explained in Ref. \cite{14}, so does $\theta_{13}$. However, the reverse statement is not true. One could have a vanishing leptonic CP asymmetry simply by setting $b_{1,2}$ to zero in  (\ref{epslnsn}). But, so long as Im $m_D^{CES}$ is nonzero, e.g. through nonvanishing values of $c_2$ and $d_2$, $\theta_{13}$ need not vanish. Indeed, the leptonic CP asymmetry depends rather sensitively on $b_{1,2}$. We shall elaborate on this later in our numerical discussion.

\section{Boltzmann equations and baryon asymmetry in different mass regimes}\label{s4}
The Boltzmann equations of concern to us govern the evolution of the number densities of the hierarchical heavy neutrinos $N_i$ and the left chiral lepton doublets $\slashed{L}_\alpha$. We  follow here the treatment given in Ref.\cite{21}. The equations involve decay transitions between $N_i$ and $\slashed{L}_\alpha \phi$ as well as $\slashed{L}_\alpha^C \phi^\dagger$ plus scattering transitions  $Q u^C\leftrightarrow N_i\slashed{L}_\alpha,\slashed{L}_\alpha Q^C\leftrightarrow N_i u^C,\slashed{L}_\alpha u \leftrightarrow N_i Q,\slashed{L}_\alpha \phi \leftrightarrow N_i V_\mu, \phi^\dagger V_\mu\leftrightarrow N_i\slashed{L}_\alpha,\slashed{L}_\alpha V_\mu \leftrightarrow N_i \phi^\dagger $. Here $Q$ represents the left-chiral quark doublet with $Q^T=(u_L\hspace{2mm}d_L)$ and $V_\mu$ can stand for either $B$ or $W_{1,2,3}$. We had already introduced in Sec. \ref{s1} the variable $z=M_{1}/T$ and the parametric function $\eta_a(z)$. When in thermal equilibrium, the latter is denoted by $\eta^{eq}_a(z)$. Recall that the number density of a particle of species $a$ and mass $m_a$ with $g_a$ internal degrees of freedom is given by\cite{22} 
\bea
n_a (T)=  \frac{g_a\, m^2_a\,T\ e^{\mu_a (T)/T}}{2\pi^2}\
K_2\bigg(\frac{m_a}{T}\bigg)\;,
\eea
 $K_2$ being the modified Bessel function of the second kind with order 2. The corresponding equilibrium density, as given by setting the chemical potential $\mu_a(T)$ equal to zero, is 
\bea
n^{\rm eq}_a (T)=  \frac{g_a\, m^2_a\,T\ }{2\pi^2}\
K_2\bigg(\frac{m_a}{T}\bigg).
\eea

We are now in a position to make use of the Boltzmann evolution equations given in Ref.\cite{21} -- generalized with flavor\cite{add1}. In making this generalization, one comes across a subtlety: the active flavor in the mass regime (given by the  value of $M_1$) under consideration may not be individually $e$, $\mu$ or $\tau$ but some combination thereof. So we use a general flavor index $\lambda$ for the lepton asymmetry. Now we write
\begin{eqnarray}
  \label{BEN} 
\frac{d \eta_{N_i}}{dz} &=& \frac{z}{H(z=1)}\ \bigg[\,\bigg( 1
\: -\: \frac{\eta_{N_i}}{\eta^{\rm eq}_{N_i}}\,\bigg)\,\sum\limits_{\beta=e,\mu,\tau} \bigg(\,
\Gamma^{\beta Di} \: +\: \Gamma^{\beta Si}_{\rm Yukawa}\: +\:
\Gamma^{\beta Si}_{\rm Gauge}\, \bigg)\nonumber\\ &&-\frac{1}{4}\sum\limits_{\beta=e,\mu,\tau}\eta_L^\beta\varepsilon^\beta_i \bigg(\,
\Gamma^{\beta Di} \: +\: \tilde{\Gamma}^{\beta Si}_{\rm Yukawa}\: +\:
\tilde{\Gamma}^{\beta Si}_{\rm Gauge}\ \bigg) \bigg],\nonumber\\
\frac{d \eta^\lambda_L}{dz} &=& -\, \frac{z}{H(z=1)}\, \bigg [\,
\sum\limits_{i=1}^3\,\varepsilon^\lambda_i \ 
\bigg( 1 \: -\: \frac{\eta_{N_i}}{\eta^{\rm eq}_{N_i}}\,\bigg)\,\sum\limits_{\beta=e,\mu,\tau} \bigg(\,
\Gamma^{\beta Di} \: +\: \Gamma^{\beta S i}_{\rm Yukawa}\: +\:
\Gamma^{\beta S i}_{\rm Gauge}\, \bigg) \nonumber\\ 
&&+\, \frac{1}{4}\, \eta^\lambda_L\, \bigg \{\, \sum\limits_{i=1}^3\, 
\bigg(\, \Gamma^{\lambda D i} \: +\: 
\Gamma^{\lambda Wi}_{\rm Yukawa}\: 
+\: \Gamma^{\lambda Wi}_{\rm Gauge}\,\bigg)\: +\:
\Gamma^{\lambda \Delta L =2}_{{\rm Yukawa}} \bigg \}\,\bigg ]\,.\label{BEL}
\end{eqnarray} 
In each RHS of (\ref{BEL}), apart from the Hubble rate of expansion $H$ at the decay temperature, we have various transition widths $\Gamma$ originally introduced in Ref.\cite{22} which are linear combinations (normalized to the photon density) of different CP conserving collision terms $\gamma_Y^X$  for the transitions $X\rightarrow Y$ and $\bar{X}\rightarrow \bar{Y}$. Here $\gamma^X_Y$ is defined as
\begin{equation}
  \label{CT}
\gamma^X_Y\ \equiv \ \gamma ( X\to Y)\: +\: \gamma ( \overline{X}
\to \overline{Y} )\;,
\end{equation}
with 
\begin{equation}
\gamma ( X\to Y)\ =\ \int\! d\pi_X\, d\pi_Y\, (2\pi )^4\,
\delta^{(4)} ( p_X - p_Y )\ e^{-p^0_X/T}\, |{\cal M}( X \to Y )|^2\; . \label{gmint}
\end{equation}
In (\ref{gmint}) one has used a short hand notation for the phase space 
\bea
d\pi_x=\frac{1}{S_x} \prod\limits_{i=1}^{n_x} \frac{d^4 p_i}{(2\pi)^3}\delta(p_i^2-m_i^2)\theta(p_i^0)
\eea
with $S_X =n_{id}!$ being a symmetry factor in case the initial state $X$ contains a number $n_{id}$ of identical particles. Moreover, the squared matrix element in  (\ref{gmint}) is summed  (not averaged) over the internal degrees of freedom of the initial and final states.\\ 

The transition widths $\Gamma$ in (\ref{BEL}) are given as follows:
\begin{eqnarray}
  \label{GD}
&&\Gamma^{\lambda Di}  =  \frac{1}{n_\gamma}\ \gamma^{N_i}_{\slashed{L}_\lambda \phi^{\dagger}}\;, \\
  \label{GSY}
&&\Gamma^{\lambda Si}_{\rm Yukawa}  = \frac{1}{n_\gamma}\
\bigg(\, \gamma^{N_i \slashed{L}_\lambda}_{Q u^C}\: +\:  \gamma^{N_i u^C}_{\slashed{L}_\lambda Q^C}\: 
+\: \gamma^{N_i Q}_{\slashed{L}_\lambda u}\, \bigg)\; ,\\
&&\widetilde{\Gamma}^{\lambda Si}_{\rm Yukawa} = \frac{1}{n_\gamma}\
\bigg(\, \frac{\eta_{N_i}}{\eta^{\rm eq}_{N_i}}\, \gamma^{N_i \slashed{L}_\lambda}_{Q u^C}\: 
+\: \gamma^{N_i u^C}_{\slashed{L}_\lambda Q^C}\: +\: \gamma^{N_i Q}_{\slashed{L}_\lambda u}\, \bigg)\;,\\
  \label{GSG}
&&\Gamma^{\lambda Si}_{\rm Gauge}  =  \frac{1}{n_\gamma}\ 
\bigg(\, \gamma^{N_i V_\mu}_{\slashed{L}_\lambda ~\phi}\: +\: 
\gamma^{N_i \slashed{L}_\lambda}_{\phi^\dagger V_\mu}\: +\: 
\gamma^{N_i\phi^\dagger }_{\slashed{L}_\lambda V_\mu}\, \bigg)\;,\\
&&\widetilde{\Gamma}^{\lambda Si}_{\rm Gauge} = \frac{1}{n_\gamma}\ 
\bigg(\, \gamma^{N_i V_\mu}_{\slashed{L}_\lambda \phi}\: +\: 
\frac{\eta_{N_i}}{\eta^{\rm eq}_{N_i}}\, 
\gamma^{N_i \slashed{L}_\lambda}_{\phi^\dagger V_\mu}\: +\: 
\gamma^{N_i\phi^\dagger }_{\slashed{L}_\lambda V_\mu}\, \bigg)\; ,\\
  \label{GWY}
&&\Gamma^{\lambda Wi}_{\rm Yukawa}  =  \frac{2}{n_\gamma}\
\bigg(\, \gamma^{N_i\slashed{L}_\lambda}_{Q u^C}\: +\:  \gamma^{N_i u^C}_{\slashed{L}_\lambda Q^C}\: 
+\: \gamma^{N_i Q}_{\slashed{L}_\lambda u}\: +\: \frac{\eta_{N_i}}{2\eta^{\rm eq}_{N_i}}\,
\gamma^{N_i \slashed{L}_\lambda}_{Q u^C}\, \bigg)\; ,\\
  \label{GWG}
&&\Gamma^{\lambda Wi}_{\rm Gauge}  =  \frac{2}{n_\gamma}\ 
\bigg(\, \gamma^{N_i V_\mu}_{\slashed{L}_\lambda \phi}\: +\: 
\gamma^{N_i \slashed{L}_\lambda}_{\phi^\dagger V_\mu}\: +\: 
\gamma^{N_i\phi^\dagger }_{\slashed{L}_\lambda V_\mu}\: +\:
\frac{\eta_{N_i}}{2\eta^{\rm eq}_{N_i}}\, 
\gamma^{N_i \slashed{L}_\lambda}_{\phi^\dagger V_\mu}\, \bigg)\;,\\
  \label{GDL2}
&&\Gamma^{\lambda \Delta L =2}_{\rm Yukawa} = \frac{2}{n_\gamma}\sum_{\beta=e,\mu\tau}\ 
\bigg(\, \gamma^{\,\prime \slashed{L}_\lambda\phi}_{\,L^{ C}_\beta\phi^\dagger} 
+\:  2\gamma^{\slashed{L}_\lambda \slashed{L}_\beta}_{\phi^\dagger\phi^\dagger}\, \bigg)\;. 
\end{eqnarray}
The explicit expressions for  $\gamma$ and $\gamma^{\prime}$  are given in Appendix B of Ref.\cite{21}. The subscripts $D$, $S$ and $W$ stand for decay, scattering and washout respectively. We rewrite the Boltzmann equations in terms of $Y_{N_i}(z)=\eta_{N_i}(z)s^{-1}$ and certain $D$-functions of $z$ that are defined below.\\

 Consider the first  equation in (\ref{BEL}) to start with. Its second RHS term has been neglected for our assumed hierarchical leptogenesis since both  $\eta_L^\beta$ and $\varepsilon_i^\beta$ are each quite small and their product much smaller\footnote{In order of magnitude this product is $ 10^{-6}\times10^{-5}\sim 10^{-11}$, as compared with the first term which is $\mathcal{O}(1)$.}. Using some shorthand notation, as explained in Eqs. (\ref{ki1}) - (\ref{ki2}) below, we can now write 
\begin{equation}
\frac{d Y_{N_i(z)}}{d z}=\{D_i(z)+D^{\rm SY}_i(z)+D^{\rm SG}_i(z)\}\{(Y^{\rm eq}_{N_i}(z)-Y_{N_i}(z)\}\label{BEN_Y1},
\end{equation}
where
\begin{eqnarray}
&&D_i(z)=\sum \limits_{\beta=e,\mu,\tau} D^\beta_i(z) = \sum \limits_{\beta=e,\mu,\tau} \frac{z}{H(z=1)}\frac{\Gamma^{\beta Di}}{\eta^{\rm eq}_{N_i}(z)},\label{ki1}\\
&&D^{\rm SY}_i(z)=\sum \limits_{\beta=e,\mu,\tau} \frac{z}{H(z=1)}\frac{\Gamma^{\beta S i}_{\rm Yukawa}}{\eta^{\rm eq}_{N_i}(z)},\\
&&D^{\rm SG}_i(z)=\sum \limits_{\beta=e,\mu,\tau} \frac{z}{H(z=1)}\frac{\Gamma^{\beta S i}_{\rm Gauge}}{\eta^{\rm eq}_{N_i}(z)}.\label{ki2}
\end{eqnarray}
Turning to the second equation in (\ref{BEL}) and neglecting the $\Delta L=2$ scattering terms, we rewrite it as
\begin{eqnarray}
\frac{d \eta^\lambda_L(z)}{dz} = & -& \sum\limits_{i=1}^3\,\varepsilon^\lambda_i \{D_i(z)+D^{\rm SY}_i(z)+D^{\rm SG}_i(z))(\eta^{\rm eq}_{N_i}(z)-\eta_{N_i}(z)\}
\nonumber\\& - & \frac{1}{4}\eta^\lambda_L\sum\limits_{i=1}^3 
\{\frac{1}{2}D^\lambda_i(z)z^2 K_2(z)+D^{\lambda \rm YW}_i(z)+ D^{\lambda \rm GW}_i(z) )\}  \label{BEL1}
\end{eqnarray}
with
\begin{eqnarray}
&&D^{\rm YW}_i(z)=\sum \limits_{\beta=e,\mu,\tau} \frac{z}{H(z=1)}\Gamma^{\beta Wi}_{\rm Yukawa},\\
&&D^{\rm GW}_i(z)=\sum \limits_{\beta=e,\mu,\tau} \frac{z}{H(z=1)}\Gamma^{\beta Wi}_{\rm Gauge}.
\end{eqnarray}
\\
A major simplification (\ref{BEL1}) occurs in our model when the active flavor $\lambda$ equals $e$ since $\varepsilon_1^e=0$ and only the second RHS term contributes to the evolution of $\eta^\lambda$. Then the solution of the equation 
becomes \cite{23}
\bea
\eta_L^e(z)=\eta_L^e(z=0)\exp[{-\frac{1}{4}\int_{0}^{z}W^e(z^\prime)dz^\prime }],\label{ete}
\eea
where 
\bea
W^e(z)=
\frac{1}{2}D^e_1(z)z^2 K_2(z)+D^{e \rm YW}_1(z)+ D^{e \rm GW}_1(z)  .
\eea
However, at a very high temperature, the lepton asymmetries get efficiently washed out. Therefore $\eta_L^e(z\rightarrow0)$ vanishes and from (\ref{ete}) $\eta_L^e(z)=0$ for all $z$. Similarly, for an unflavored (i.e flavor-summed) leptogenesis in our model, $\eta^e+\eta^\mu+\eta^\tau=0$ since $\varepsilon_1^\mu=-\varepsilon_1^\tau$.\\

We are now ready to calculate the baryon asymmetry from the lepton asymmetry.  To that end, it is first convenient to define the variable 
\bea
Y_\lambda=\frac{n^\lambda_L-n^\lambda_{\bar{L}}}{s}=\frac{n_\gamma}{s}\eta_L^\lambda,
\eea
i.e. the leptonic minus the antileptonic number density of the active flavor $\lambda$ normalized to the entropy density. The factor  $s/\eta_\gamma$ is known  to equal $1.8g_{\ast s}$ and is a function of temperature. For $T>10^2~{\rm GeV}$, $g_{\ast s}$ is known to remain nearly constant with temperature at a value 
(with three right chiral neutrinos) of about $112$\cite{24}. Sphaleronic processes convert the lepton asymmetry created by the decay of the right chiral heavy neutrinos into a baryon asymmetry by keeping
 $\Delta_\lambda=\frac{1}{3}B-L^\lambda$ conserved.  $Y_{\Delta_\lambda}$, defined as $s^{-1}\{1/3(n_B-n_{\bar{B}})-(n_L-n_{\bar{L}})\}$, and $Y_\lambda$  are linearly related, as under
 \bea 
Y_\lambda=\sum\limits_{\rho}A_{\lambda\rho}Y_{\Delta_\rho}, \label{yalph}
\eea
where $A_{\lambda\rho}$ are a set of numbers whose values depend on which of the three mass regimes in which $M_{1}$ lies, as mentioned in the Introduction. These are discussed in detail later in the section. Meanwhile, we can rewrite (\ref{BEL1}) as
\begin{eqnarray}
\frac{d Y_{\Delta_\lambda}}{dz}& = & \sum\limits_{i=1}^3[\varepsilon^\lambda_i \{D_i(z)+D^{\rm SY}_i(z)+D^{\rm SG}_i(z)\}\{Y^{\rm eq}_{N_i}(z)-Y_{N_i}(z)
\} ] \nonumber\\ & + & \frac{1}{4}\sum\limits_{\rho}A_{\lambda\rho}Y_{\Delta_\rho} \sum\limits_{i=1}^3 
\{\frac{1}{2}D^\lambda_i(z)z^2 K_2(z)+D^{\lambda ~\rm YW}_i(z)+ D^{\lambda ~\rm GW}_i(z)\}. \label{BEL_Y1}
\end{eqnarray}
 We need to solve (\ref{BEN_Y1}) and (\ref{BEL_Y1}) and evolve $Y_{N_i}$ as well as $Y_{\Delta_\lambda}$ upto a value of $z$ where the quantities
$Y_{\Delta_\lambda}$ become constant with $z$, i.e. do not change as $z$ is varied. The final baryon asymmetry $Y_B$ is obtained \cite{25} linearly in terms $Y_{\Delta_\lambda}$, the coefficient depending on the mass regime in which $M_{1}$ is located, as explained in what follows. Let us then discuss three mass regimes separately.
\subsection{${\bf{M_{1}<{10}^{9}}}$ GeV}\label{ff}
Here all three lepton flavors are separately distinguishable. Therefore the flavor index  $\lambda$ can just be $\lambda=e$ or $\mu$ or $\tau$. In this case the $3\times3$ $A$ matrix, whose $\lambda,\rho$ element relates $Y_\lambda$ and $Y_{\Delta_\rho}$, is given by\cite{7}
\begin{equation}
A=\left(\begin{array}{ccc}
-151/179 & 20/179 & 20/179\\
25/358 & -344/537 & 14/537\\
25/358 &  14/537 &  -344/537
\end{array}\right).
\end{equation}
Now the final baryon asymmetry  normalized to the entropy density, is given by\cite{26}
\begin{equation}
Y_B=\frac{28}{79}( Y_{\Delta_\mu}+Y_{\Delta_\tau}),  \label{ff_yb}
\end{equation}
since $Y_{\Delta_e}$ vanishes on account of $\eta_L^e$ being zero.
Another important parameter, namely the  baryon asymmetry normalized to the photon
density,  obtains  as
\begin{equation}
\eta_B=\left.\frac{s}{n_\gamma}\right|_0Y_B=7.0394Y_B,
\end{equation}
 the subscript zero denoting the present epoch.

\subsection{ ${\bf{{10}^{9}\,\,{\rm{\bf{{\rm GeV}}}}<M_{1}<{10}^{12}}}$ GeV}\label{pf}
In this regime the $\tau$ flavor is distinguishable but one cannot
differentiate between the $e$ and $\mu$ flavors. It is therefore convenient to define two sets of CP asymmetry parameters  $\varepsilon^\tau$ and 
$\varepsilon^{(2)}=\varepsilon^e+\varepsilon^\mu$. Therefor the index $\lambda$ takes the values $\tau$ and $2$. The Boltzmann equations lead to the two asymmetries $Y_{\Delta_\tau}$ and $Y_{\Delta_2}$. These are related to $Y_\tau$ and $Y_2=Y_e+Y_\mu$  by a $2\times2$ A-matrix  given by \cite{7}
\begin{equation}
A=\left(\begin{array}{cc}
  -417/589  & 120/589\\
   30/589 & -390/589
\end{array}\right). 
\end{equation}
The final baryon asymmetry $Y_B$ is then calculated as\cite{7}
\begin{equation}
Y_B=\frac{28}{79}(Y_{\Delta_2}+Y_{\Delta_\tau}) \label{pf_yb}.
\end{equation}
\subsection{ ${\bf{M_{1}>{10}^{12}}}$ GeV}\label{uf}
In this case all the lepton flavors act indistinguishably leading to a single CP asymmetry parameter $\varepsilon_i=\sum\limits_{\lambda}\varepsilon^\lambda_i$. As mentioned earlier, $\sum\limits_{\lambda}\eta_L^\lambda=0$ and $Y_{\Delta}=0$. Therefore $Y_B=0$ and no baryogenesis is possible in this mass regime. This statement is independent of the mass ordering of the light neutrinos.

\section{Numerical analysis: methodology and discussion} \label{s5}
\begin{figure}
\begin{center}
\includegraphics[scale=0.4]{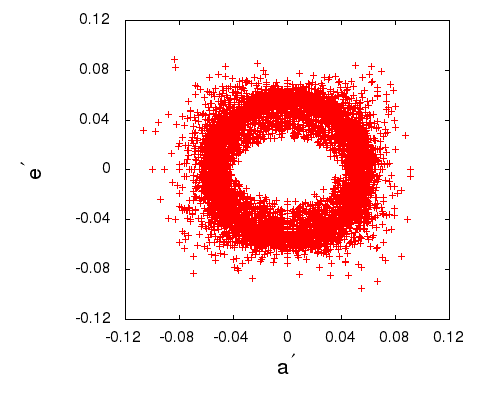}\includegraphics[scale=0.4]{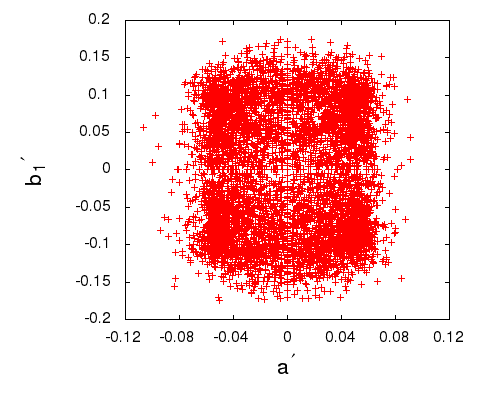}\\
\includegraphics[scale=0.4]{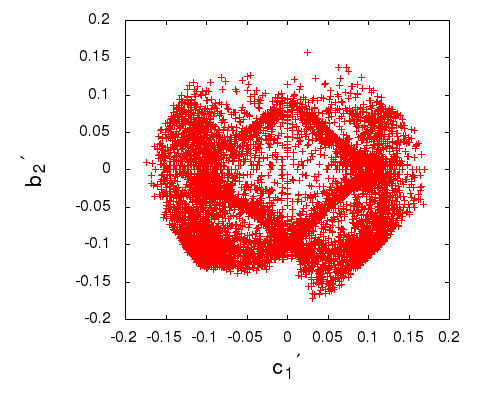}\includegraphics[scale=0.4]{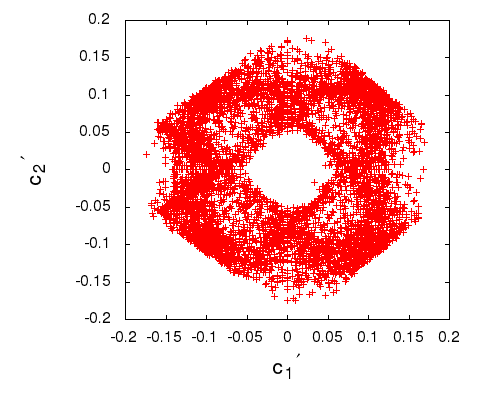}
\caption{ Plots of the reduced parameters for a normal mass  ordering of the light neutrinos.}\label{f2}
\end{center}
\end{figure}

\begin{figure}
\begin{center}
\includegraphics[scale=0.4]{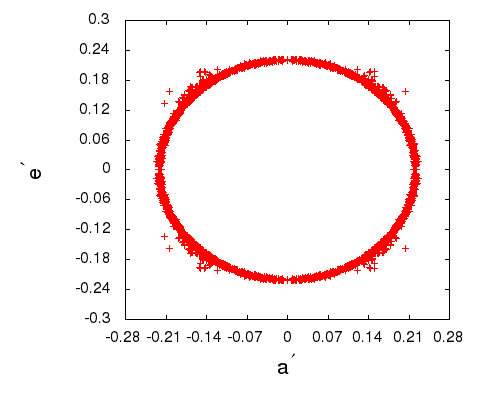}\includegraphics[scale=0.4]{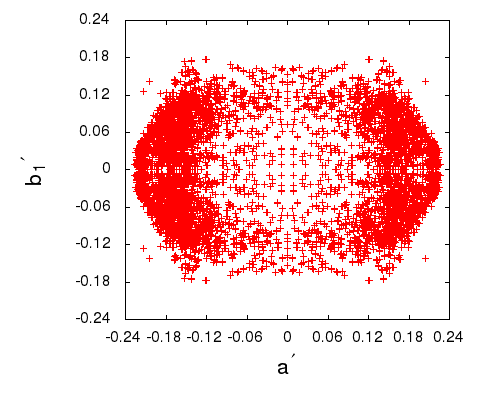}\\
\includegraphics[scale=0.4]{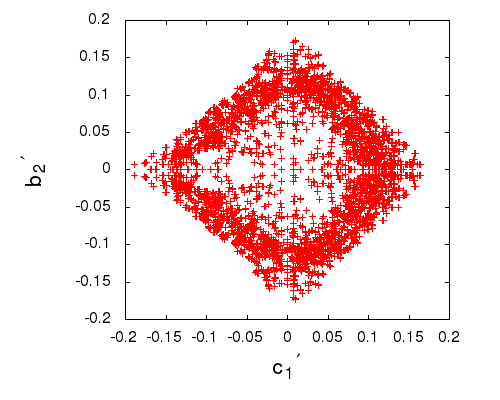}\includegraphics[scale=0.4]{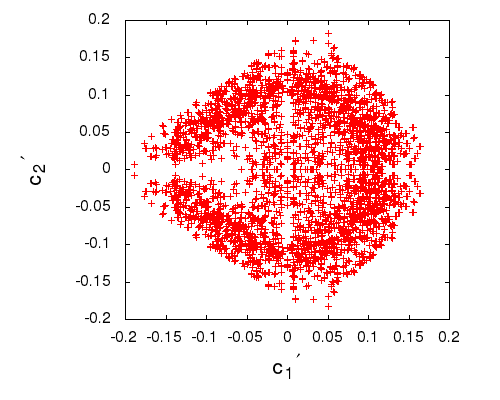}
\caption{Plots of the reduced parameters for an inverted mass  ordering of the light neutrinos.}\label{f3}
\end{center}
\end{figure}

\begin{figure}[H]
\begin{center}
\includegraphics[scale=0.4]{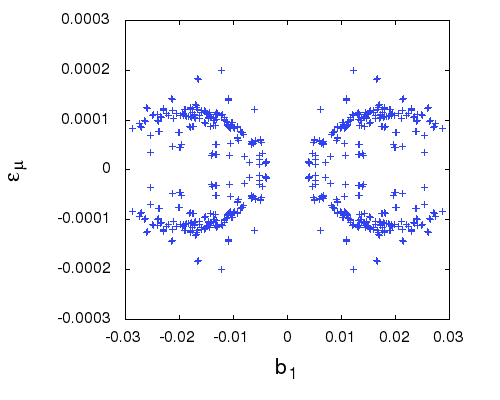}\includegraphics[scale=0.4]{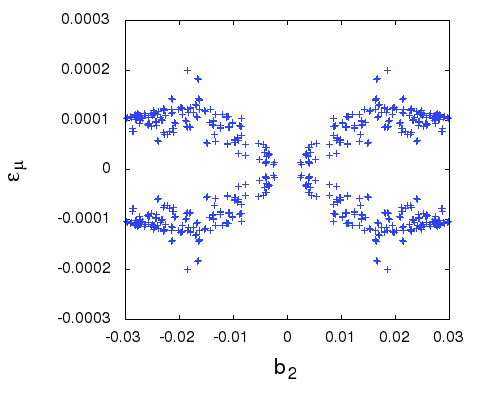}\\
\caption{Plot of $\varepsilon^{\mu}_1$ with $b_1$ (left), $b_2$ (right) for a normal light neutrino mass  ordering.}\label{f4}
\end{center}
\end{figure}

In order to numerically check the viability of our theoretical results, the allowed ($3\sigma$) values of globally fitted neutrino  oscillation  data \cite{27} and the upper bound of 0.23 eV on the sum of the light neutrino masses  have been used, cf. Table \ref{osc1}. We  first constrain the  parameter space constructed with the six rescaled parameters defined in Eqs. (\ref{rsc1}) - (\ref{rsc2}). Both  normal and inverted types of light neutrino mass ordering are found to be allowed over a sizable  region  of the parameter space consistent with the input constraints.  The ranges  of 
\begin{table}[H]
\begin{center}
\caption{Input values used} \label{osc1}
 \begin{tabular}{|c|c|c|c|c|c|c|} 
\hline 
${\rm Parameters}$&$\theta_{12}$&$\theta_{23}$ &$\theta_{13}$ &$ \Delta m_{21}^2$&$|\Delta m_{31}^2|$&$\Sigma_i m_i$ \\
&$\rm degrees$&$\rm degrees$ &$\rm degrees$ &$ 10^{-5}\rm eV^2$&$10^{-3} \rm (eV^2)$&$ \rm (eV) $ \\
\hline
$3\sigma\hspace{1mm}{\rm ranges/\hspace{1mm}others\hspace{1mm}}$&$31.29-35.91$&$38.3-53.3$&$7.87-9.11$&$7.02-8.09$&$2.32-2.59$&$<0.23$\\
\hline
${\rm Best\hspace{1mm}{\rm fit\hspace{1mm}}values\hspace{1mm}(NO)}$&$33.48$&$42.3$&$8.50$&$7.50$&$2.46$&$-$\\
\hline
${\rm Best\hspace{1mm}{\rm fit\hspace{1mm}}values\hspace{1mm}(IO)}$&$33.48$&$49.5$&$8.51$&$7.50$&$2.45$&$-$\\
\hline
\end{tabular} 
\end{center} 
\end{table}

\noindent  
the rescaled parameters are graphically shown in Fig.\ref{f2} and Fig.\ref{f3} respectively for the normal and the inverted  ordering of the light neutrino masses.
This is the primary constraining procedure  since the CP asymmetry parameters $\varepsilon^\alpha_{i}$  and the different $\Gamma$'s of the Boltzmann equations depend individually upon the elements of $m_D$ and the RH neutrino masses $M_i\hspace{1mm}(i=1,2,3)$. Therefore, merely restricting the rescaled parameters is not sufficient for the computation of the final baryon asymmetry. In order to obtain the allowed ranges of the parameters  $a$, $b_{1,2}$, $c_{1,2}$  and $e$, included in $m_D$, we incorporate the strong hierarchy assumption of the RH neutrino masses ($M_1<<M_2<<M_3$), as mentioned in  earlier sections. For numerical purposes, we arbitrarily choose $M_2/M_1$ = $M_3/M_2=10^{3}$. We shall later discuss in Section \ref{s6} the effects of changing these mass ratios. Depending upon the mass regime, for a fixed value of  $M_1$, we then obtain the allowed ranges of the parameters of $m_D$ from the relations defined in Eqs. (\ref{rsc1}) - (\ref{rsc2}). 
\paragraph{}
Even after constraining  the six unprimed parameters of $m_D$ and the masses of the three right handed heavy neutrinos, three  undetermined parameters remain -- namely $f$, $d_1$ and $d_2$. The latter have been neglected earlier in the primary implementation of the input constraints since their contributions  to the light neutrino mass matrix $M_\nu$ are suppressed by the heaviest RH neutrino mass $M_3$. However, for a quantitatively successful  treatment of  leptogenesis, one needs to estimate these missing parameters too, as mentioned in Sec. \ref{s2}. We discuss here some  technical details regarding this estimation.  For example, let us consider the first equation of Table \ref{t}, namely
\begin{equation}
 x=-\bigg( \frac{a^2}{M_1} + \frac{e^2}{M_2} + \frac{f^2}{M_3}\bigg). 
\end{equation}
The last RHS term was earlier neglected on the grounds that the parameter $f$, which is presumably of same the order of magnitude as $a$ or $e$, is suppressed by $M_3$. Now, in order to estimate $f$,  we first set it at a value which is larger i.e. between $a$ and $e$. Then we keep on decreasing it until the quantity $f^2M_3^{-1}/(a^2M_1^{-1}+e^2M_2^{-1})$ becomes less than a very small number which we choose to be $10^{-5}$. In a similar manner one can estimate approximate values of $d_1$ and $d_2$. Thus, knowing  the numerical values of all the parameters of $m_D$
as well as those of $M_R$, we can make a realistic estimate of  the final value of the baryon asymmetry. The first step towards the last-mentioned goal is the estimation of $\varepsilon^{\lambda}_1$  in the three mass regimes of $M_1$.
 We have carried out our  numerical analysis over a wide range of values of $M_1$ in the  $\tau$-flavored and in the fully flavored regimes.  As  mentioned in the last paragraph of Sec. \ref{s3},  $\varepsilon^{\mu,\tau}_1$ are mostly sensitive to $b_{1,2}$. In order to see the nature of the variation of  $\varepsilon^{\mu,\tau}_1$ with $b_{1,2}$ for  constant values of $c_2$ and $d_2$, we first set $c_2$ and $d_2$  to be zero. Now the simplified expression of the relevant CP asymmetry parameter becomes
\begin{eqnarray}
&& \varepsilon^\mu_1= \xi(b_2k^2\chi_2 + b_1 \chi_3^\prime)=-\varepsilon^\tau_1,
\end{eqnarray} 
where $\xi$ and $\chi_2$ as are defined in (\ref{xi}), and $\chi_3^\prime$  is given by
\bea
\chi_3^\prime = b_2(1+k^2) [ c_1^2\{1+g(x_{12})-x_{12})+d_1^2(1+g(x_{13})-x_{13}\}].
\eea 
For a graphical representation of the variation of the CP asymmetry parameter $\varepsilon^{\mu}_1$  with $b_{1,2}$, we choose a sample value of $M_1=$ $3.62\times10^{11}$ GeV and assume a normal mass ordering\footnote{ As we shall see later, in our model an inverted mass ordering is disfavored in terms of a realistic baryogenesis.} of the light neutrinos. The corresponding scatter plots are shown in Fig. \ref{f4}. The vanishing  of $b_{1,2}$ implies $\varepsilon^{\mu}_1=0$; therefore, in our  

\noindent
 numerical computation, only those values of $\varepsilon^{\mu}_1$ are allowed which correspond to $b_{1,2}\neq 0$. One can have a similar plot for $\varepsilon^{\tau}_1$  since $\varepsilon^{\mu}_1=-\varepsilon^{\tau}_1$ and the plots in Fig. \ref{f4} are symmetric about the origin. The corresponding plots for an inverted mass ordering of the light neutrinos can also be generated. However, with the same computational technique as used for normal ordering, we find a much smaller number of allowed points which hardly show a fair variation of $\varepsilon^{\mu}_1$ with $b_{1,2}$. 

\paragraph{}
Finally, knowing the numerical range of $\varepsilon^{\lambda}_1$ is the last step needed to solve the Boltzmann equations  given in (\ref{BEN_Y1}) and (\ref{BEL_Y1}) leading to the parameter  $Y_{\Delta_\lambda}$ upto a fairly large value of $z$ where $Y_{\Delta_\lambda}$ becomes constant.  Then, using the suitable equations (\ref{ff_yb}), (\ref{pf_yb}),  depending upon the energy regime, one can compute the final value of $Y_B$. However,  this final step needs to overcome the following hurdle. Unlike estimating $\varepsilon^{\lambda}_1$ for the entire allowed parameter ranges of $m_D$ and $M_R$, it becomes impractical in terms of computer time to solve the Boltzmann equations for this huge  data set even if $M_1$ is fixed to a constant value. So we were obliged to use only those values of the members of parameter set for which the neutrino oscillation observables are restricted close to their best fit values.  For this purpose we choose a $\chi^2$ for every observable deviating  from its experimentally
measured best fit value as
\begin{equation}
\chi^2= \sum\limits_{i=1}^5 \Big[ \frac{O_i(th)-O_i(bf)}{\Delta O_i} \Big]. \label{chi2}
\end{equation}
In (\ref{chi2}) $O_i$ denotes the $i$th neutrino oscillation observable from among $(\Delta m^2_{21},\Delta m^2_{32},\theta_{12},\theta_{23},\theta_{13})$
and the summation runs over all the five observables. The parenthetical  $th$ stands
for the theoretical prediction, i.e the numerical value of the observable given by our model, whereas
 $bf$ denotes the best fit value (cf. Table \ref{osc1}). $\Delta O_i$ in the denominator stands for the measured $1\sigma$
range of $O_i$. After calculating $\chi^2$ for all the points $\{a^\prime, e^\prime,b_1^\prime,c_1^\prime,b_2^\prime,c_2^\prime\}$, as 
allowed by the oscillation data, we start from the minimum value of the $\chi^2$ (= $\chi^2_{min}$) and keep on increasing the latter until we get $Y_B$ to be positive as well as in the observed range.  It is to be noted that for a particular value of $\chi^2$, i.e. for a particular primed data set,  we are able to generate a large number of unprimed points (parameters of $m_D$) by varying the values of $M_1$ in  Eqs. (\ref{rsc1})-(\ref{rsc2}). To be more precise,  \textquoteleft$n$\textquoteright ~values of $M_1$ lead to \textquoteleft$n$\textquoteright~  values of the unprimed set of parameters for the particular primed set under consideration.  The other three parameters $f$, $d_1$ and $d_2$ are again computed by means of the previously mentioned approximation technique. We vary $M_1$ over a wide range in  the relevant mass regimes for both  types of mass ordering and present our final result systematically in the following way.\\

\noindent
\textbf{$Y_B$ for normal mass ordering of light neutrinos:}\\

\paragraph{$\bf{M_{1}<{10}^{9}}$ GeV:} In  this regime all  lepton flavors $(e,\mu,\tau)$ act distinguishably. However, since $\varepsilon_1^{e}=0$, we first need to evaluate $\varepsilon_1^{\mu,\tau}$ individually. It is found that $|\varepsilon_1^{\mu,\tau}|$ can have values at most $\sim 10^{-8}$.  $Y_B$ of the right amount cannot be generated  with such a small CP asymmetry parameter\cite{5}.  

\begin{center}
\begin{figure}[!ht]
\includegraphics[width=5.5cm,height=5.5cm,angle=0]{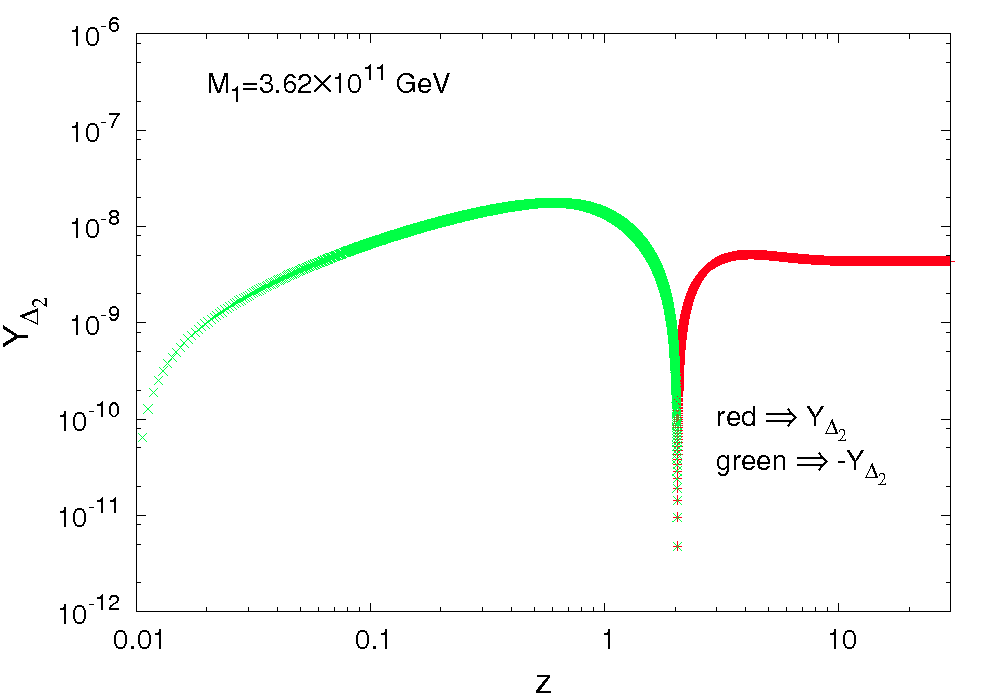}
\includegraphics[width=5.5cm,height=5.5cm,angle=0]{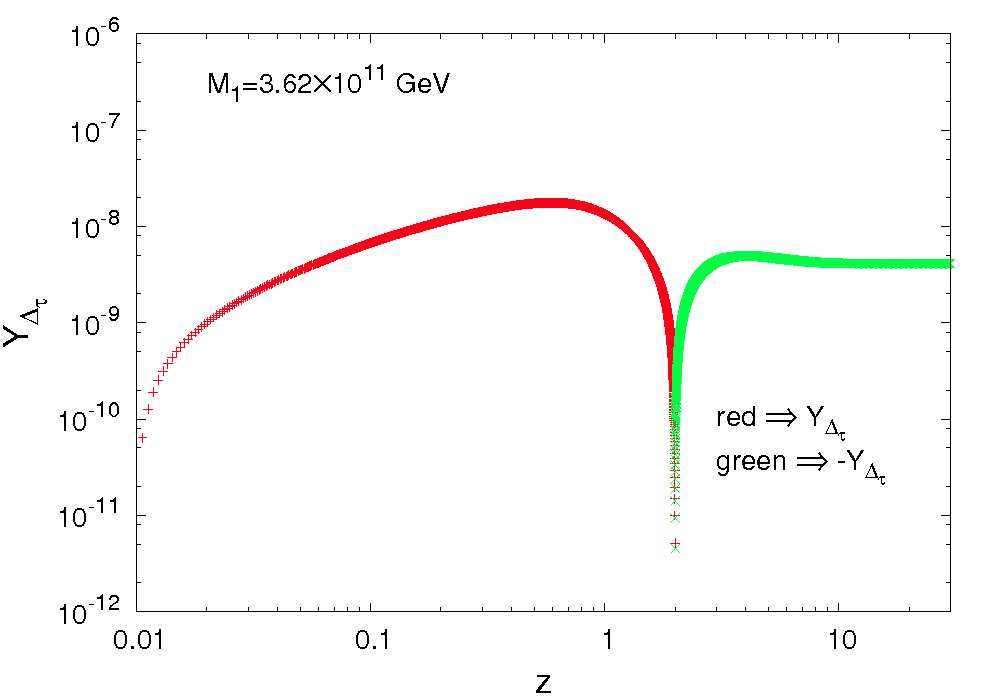}
\includegraphics[width=5.5cm,height=5.5cm,angle=0]{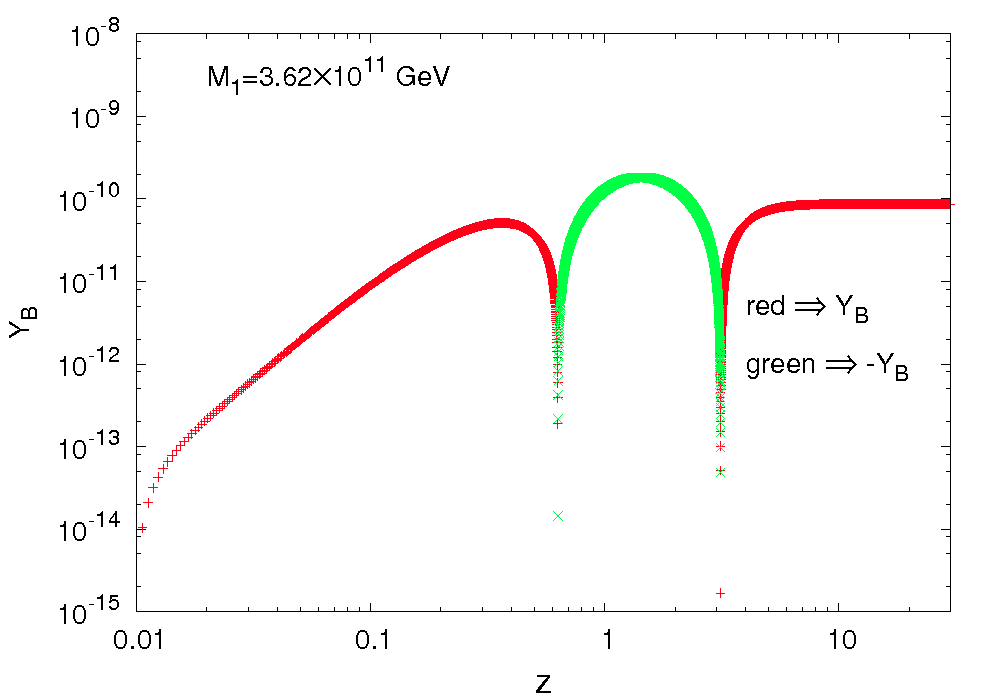}
\caption{Variation of $Y_{\Delta_\mu}$ (left), $Y_{\Delta_\tau}$ (middle), $Y_B$ (right) with $z$ in the mass regime (2) for a definite value of $M_1$. N.B. since these become negative for certain values of $z$, their negatives have been plotted on the log scale for those values of $z$. A normal mass ordering for the light neutrinos has been assumed.}
\label{asy_m}
\end{figure}
\end{center}

\begin{figure}[!ht]
\begin{center}
\includegraphics[scale=.45]{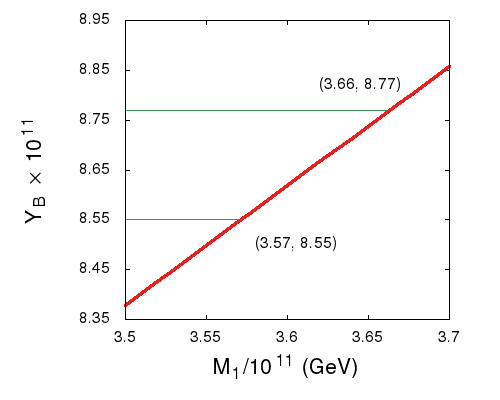}
\caption{A plot of the final $Y_B$ for different values of $M_1$ for a normal light neutrino mass ordering.}
\label{yb_m1}
\end{center}
\end{figure}

\noindent
\paragraph{$\bf{{10}^{9}\,\,{\rm{\bf{{\rm GeV}}}}<M_{1}<{10}^{12}}$ GeV:} After carrying out the $\chi^2$ analysis for this regime, we first calculate the final $Y_B$ for $\chi^2_{min}(=0.002)$. It is found that the final $Y_B$ saturates to a negative value. Then we keep on increasing $\chi^2$ and find that   a positive value for the final $Y_B$ within the observed range may be obtained for $\chi^2=0.003$ which is close enough to the best-fit value of $\chi^2=0.002$. In the entire analysis, for each value of $\chi^2$, i.e. for this single primed set, $M_1$  is varied over a wide range. Then, for each value of $M_1$, a set of values of the unprimed parameters $\{a,e,f,b_1,c_1,d_1,b_2,c_2,d_2 \}$ is generated. The Boltzmann equations are solved for each set of values of $M_1$. 
Since, in this regime, the  $\tau$ flavor acts distinguishably,  we need  to solve
\begin{table}[H]
\caption{parameters corresponding  $\chi^2=0.003$ for normal mass ordering.}
\label{chi_nh}
\begin{center}
 \begin{tabular}{ |c|c|c|c|c|c|c| } 
 \hline
 $a^\prime$ & $e^\prime$ & $b_1^\prime$ & $c_1^\prime$ & $b_2^\prime$& $c_2^\prime$& $\chi^2$ \\ \hline
 $0.026$ & $0.054$ & $0.019$ & $0.095$& $-0.080$& $0.095$& $0.003$\\ \hline
\end{tabular}
\end{center}
\end{table}
\noindent
 the Boltzmann equations for two flavors ($\tau$ and 2) in order to obtain the variation of $Y_{\Delta_{\tau,2}}$ or of $Y_B$ with $z$. For each set of the primed parameters, we take thirty values of $M_1$ within the range $10^{9}$ GeV to $10^{12}$ GeV and solve the Boltzmann equations thirty times for each $M_1$ along with the corresponding unprimed set of rescaled parameters. For a concise presentation, in Table \ref{Yb_m}, we tabulate only ten such values of $M_1$  for which $Y_B$ is near or inside the observed range. 
\begin{table}[!ht] 
\caption{$Y_B$ for different masses of lightest right handed neutrino.}
\label{Yb_m}
\begin{center}
\begin{tabular}{ |c|c|c|c|c|c|c|c|c|c|c| } 
\hline
$\frac{M_1}{10^{11}}$ (GeV) &  $3.57$ & $3.58$ & $3.59$ & $3.60$ & $3.61$ & $3.62$ & $3.63$ & $3.64$ & $3.65$ & $3.66$ \\ \hline
$Y_B\times10^{11}$ & $8.55$  & $8.57$ & $8.59$& $8.61$& $8.64$& $8.66$& $8.69$& $8.71$& $8.74$& $8.77$\\ \hline
\end{tabular}
\end{center}
\end{table}
 Fig.\ref{asy_m} contains a graphical presentation of the variation of the  asymmetries $Y_{\Delta_2}$, $Y_{\Delta_\tau}$ and  $Y_B$ with $z$  for a definite value of $M_1$ which is taken to be $3.62\times10^{11}$ GeV. It may be seen that $Y_B$ is inside the observed range \cite{1} for large $z$ corresponding to the present epoch. A careful surveillance of  Table \ref{Yb_m} leads to the conclusion that we can obtain  upper and lower bounds on $M_1$ due to the constraint from the observed range of $Y_B$. One can appreciate this  fact more clearly from the plot of $Y_B$ vs. $M_1$ in Fig.\ref{yb_m1}. Two straight lines have been drawn parallel to the abscissa in Fig.\ref{yb_m1}: one at $Y_B=8.55\times10^{-11}$ and the other at $Y_B=8.77\times10^{-11}$.  The values of $M_1$, where the straight lines meet the $Y_B$ vs $z$ curve, yield the allowed lower and upper bounds on $M_1$, namely $(M_1)_{lower}=3.57\times10^{11}$ GeV and  $(M_1)_{upper}=3.66\times10^{11}$ GeV. 

\paragraph{$\bf{M_{1}>{10}^{12}}$ GeV:} It has been shown that $Y_B=0$ here for our model.\\

\noindent
\textbf{$Y_B$ for inverted mass ordering of light neutrinos:}\\

\noindent
In this case too  the numerical estimation of the baryon asymmetry parameter has been made exactly in the same 
manner as for a normal mass ordering. A final discussion for each  regime goes as follows. 

\paragraph{${\bf{M_{1}<{10}^{9}}}$ GeV:} As in the case of normal ordering, the values of $\varepsilon_1^{\mu,\tau}$ can reach up to at most the order of $10^{-8}$ which  is not adequate to let $Y_B$ come within its observed range.

\paragraph{$\bf{{10}^{9}\,\,{\rm{\bf{{\rm GeV}}}}<M_{1}<{10}^{12}}$ GeV:} In this regime we first calculate the minimum value of $\chi^2$ for the full  set of primed parameters constrained by the oscillation data. We find that for $\chi^2_{min}=0.246$ the final baryon asymmetry saturates to a negative value. As in the previous case we then keep on increasing the value of $\chi^2$ and check the final $Y_B$ by varying $M_1$ over a wide range for each value of $\chi^2$. It turns out that though $Y_B$ attains a positive value for $\chi^2=0.952$, it is below the observed range. Then, using the $\chi^2$ enhancement technique, for $Y_B$ to be in the observed range the minimum value of $\chi^2$ is found to be 1.67 which is far away from the best-fit point. The set of primed parameters for $\chi^2=1.67$ is tabulated in Table \ref{chi_ih}.
\noindent
\begin{table}[!ht]
\caption{parameters corresponding  $\chi^2=1.67$ for inverted hierarchy}
\label{chi_ih}
\begin{center}
 \begin{tabular}{ |c|c|c|c|c|c|c| } 
 \hline
 $a^\prime$ & $e^\prime$ & $b_1^\prime$ & $c_1^\prime$ & $b_2^\prime$& $c_2^\prime$& $\chi^2$ \\ \hline
 $0.15$ & $0.16$ & $-0.017$ & $-0.022$ & $0.10$ & $-0.096$& $1.67$\\ \hline
\end{tabular}
\end{center}
\end{table}  
 
 \paragraph{$\bf{M_{1}>{10}^{12}}$ GeV:} Once again, $Y_B=0$ here for the present model.\\

A compact presentation of the final conclusions regarding $Y_B$ from the numerical analysis  is given in Table \ref{tf}.
\begin{table}[h!]
\begin{center}
\caption{Final statements on $Y_B$ for different mass regimes. } \label{tf}
 \begin{tabular}{|c|c|c|c|} 
\hline 
${\rm Type}$&${M_1<10^{9}\hspace{1mm}{\rm GeV}}$&$10^{9}\hspace{1mm}{\rm GeV}<M_1<10^{12}\hspace{1mm}{\rm GeV}$ &${M_1>10^{12}\hspace{1mm}{\rm GeV}}$\\
\hline
$\pbox{20cm}{{\rm Normal}\\ {\rm Ordering}}$ & $\pbox{20cm}{{\rm Ruled~out~since~$Y_B$ }\\{\rm is~below~the~observed~range}\\{\rm for~any~$\chi^2$.}}$ & $\pbox{20cm}{{\rm $Y_B$ within \hspace{1mm}the~ observed \hspace{1mm} range~for~$\chi^2$=0.003}\\{\rm close~to~$\chi^2_{min}=0.002$.}}$ &$\pbox{20cm}{{\rm Ruled~out}\\{\rm since $Y_B=0.$}}$ \\
\hline
$\pbox{20cm}{{\rm Inverted}\\ {\rm Ordering}}$&$\pbox{20cm}{{\rm Ruled~out~since~$Y_B$ }\\{\rm is~below~the~observed~range}\\{\rm for~any~$\chi^2$.}}$&$ \pbox{20cm}{{\rm $Y_B$ within \hspace{1mm}the ~ observed \hspace{1mm} range~for~$\chi^2$=1.67}\\{\rm far~away~from~$\chi^2_{min}=0.246$.}}$&$\pbox{20cm}{{\rm Ruled~out}\\{\rm since $Y_B=0.$}}$\\
\hline
\end{tabular} 
\end{center} 
\end{table}

We would like to make a further statement before finishing this numerical discussion . Though we had earlier  enumerated the  difficulties in numerically solving the Boltzmann equations for each data point within the entire 3$\sigma$ parameter range of $m_D$, we have been able to perform the task only for a few data points in that range. We actually find that there is no monotonic variation of $Y_B$ with the chosen data points. For example, given a normal ordering of the light neutrino masses, suppose we take the data set that corresponds to the worst fit point ($\chi^2_{max}$) and solve the Boltzmann equations for $10^9{\rm~GeV}<M_1<10^{12}$ GeV. Such a procedure yields a negative final value of  $Y_B$  contrary to  the result obtained in the $\chi^2=0.003$ case. For the other data points also,  $Y_B$ varies widely with the parameters of $m_D$   from one neutrino mass model to another  \cite{28,29,30,31}. This conclusion is true for all  mass regimes (except for $M_1>10^{12}$ GeV, where $\sum\limits_{\lambda}\varepsilon^\lambda_1=0$ and hence $Y_B$ vanishes) as well as for an inverted mass ordering of the light neutrinos. Table \ref{tf} shows that, for data points close to the best fit values, an inverted mass ordering is not favored in this model. However, {\it we cannot completely rule out this mass ordering here since such is not the case as one moves further away from the best-fit values while still remaining within the $3\sigma$ range}. There may exist certain data sets (e.g. $\chi^2=1.67$) in the allowed 3$\sigma$ ranges for which  the proper value of $Y_B$ can be generated even with an inverted light neutrino mass ordering. 
\section{Sensitivity to the heavier neutrinos}\label{s6}
In our  analysis so far, the effect of the  two heavier neutrinos ($N_2$, $N_3$) on the produced final lepton asymmetry has been neglected. We have assumed that the asymmetries produced by the decays of both of them get  washed out\cite{32}. We examine this issue in this section. Is $Y_B$ sensitive to  $N_2$ and $N_3$? There are two ways that such a sensitivity might arise: (1) directly, if the contributions to $Y_\lambda$ from $N_{2,3}$ decays do not get washed out for some reason and (2) indirectly, even if those do get washed out, a dependence of $Y_B$ on the heavier RH neutrino masses might persist through the CP asymmetry parameter $\varepsilon_1^\alpha$.\\

\noindent
{\bf Indirect effect of $N_{2,3}$:}\\

\begin{figure}[!ht]
\includegraphics[scale=.45]{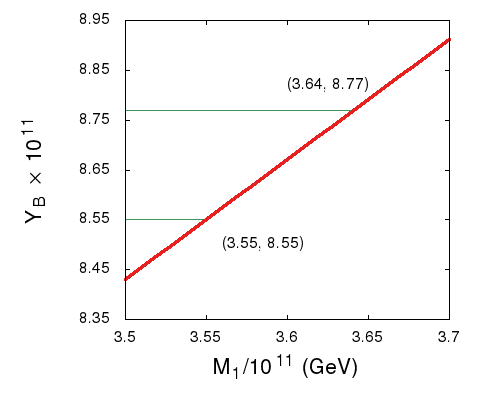} \includegraphics[scale=.45]{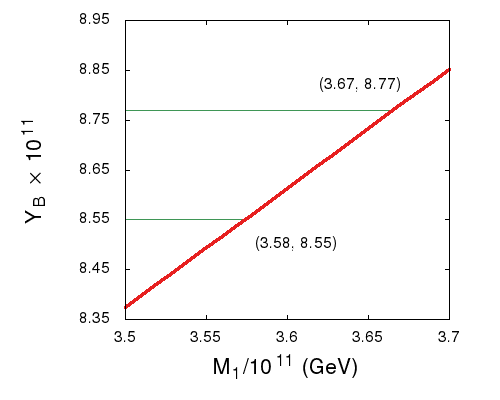}
\caption{Plots of final $Y_B$ for different values of $M_1$ for $M_{i+1}/M_i=10^2$ (left) and $M_{i+1}/M_i=10^4$ (right). }
\label{yb_m23}
\end{figure}

 Though the neutrino oscillation data have been fitted with the primed parameters, cf (\ref{rsc1})--(\ref{rsc2}), for computing  the quantities related to leptogenesis, we need to examine the  unprimed ones, i.e. the Dirac mass matrix elements. Is the final baryon asymmetry  affected by the chosen hierarchies of the RH neutrinos? Interestingly, we find that  the final $Y_B$ is not so sensitive to $M_{2,3}$.  One can justify this statement  by simplifying the CP asymmetry parameters of (\ref{epsi_intro_h}) to 
\bea
 \varepsilon_1^{\alpha}=-\frac{3}{8\pi v^2 \mathcal{H}_{11}}\sum\limits_{j =2,3} \frac{M_1}{M_j}\hspace{1mm} {\rm Im}[\hspace{1mm}\mathcal{H}_{1j}(m_D)_{1 \alpha}(m_D^*)_{j \alpha}]-\frac{1}{4\pi v^2 {\mathcal{H}}_{11}}\sum_{j=2,3} \frac{M_1^2}{M_j^2} {\rm Im}[\hspace{1mm}{\mathcal{H}}_{j1}(m_D)_{1 \alpha }({m_D}^*)_{ j \alpha }],
\label{epsim}
\eea
after approximating $g(x_{1j})$ of Eqs. (\ref{loop}) and (\ref{loop2}) to be $g(x_{1j})=-\frac{3}{2\sqrt{x_{1j}}}$ for $x_{1j}>>1$. The last term of Eq. (\ref{epsim}) is much suppressed since it is of second order in $x_{1j}^{-1}$. The first term has two parts for $j=2,3$. However, since $M_3$ is much larger than $M_1$ and  $f,d_1$ and $d_2$ are taken to have values of the order of the other Dirac components, the  $j=3$ term has a negligible effect on $\varepsilon_1^\alpha$. Now, for $j=2$, $\varepsilon_1^{\alpha}$ is simplified as 
\bea
\varepsilon_1^{\mu}=-\frac{3M_1}{8\pi v^2 \mathcal{H}_{11}}[(ae^\prime +b_1 c_1^\prime + b_2 c_2^\prime)(b_2c_1^\prime + b_1 c_2^\prime)]=-\varepsilon_1^{\tau} \label{epslnap}
\eea  
with $\varepsilon_1^{e}=0$. Since  $e^\prime$ and $c_{1,2}^\prime$ are fixed by the oscillation data, $\varepsilon_1^{\mu,\tau}$ are insensitive to the value of $M_2$. In order to numerically compute  the final baryon asymmetry for a normal mass ordering of the light neutrinos, we consider each term in (\ref{epslnap}) and two different mass hierarchical schemes for the RH neutrinos, e.g, $M_{i+1}/M_i=10^2$ and $M_{i+1}/M_i=10^4$ where $i$ can take the values 1,2. Recall that in the previous section we have presented  $Y_B$ for $M_{i+1}/M_i=10^3$. A careful inspection of Fig.\ref{yb_m1} and Fig.\ref{yb_m23} reveals an interesting fact. Though the chosen mass ratios of the RH neutrinos have been altered, changes in the lower and upper bounds on $M_1$ are not  significant for the observed range of $Y_B$. For convenience, we present in Table \ref{t8} the variation of $Y_B$ with $M_1$ for different mass ratios.

\begin{table}[h!]
\begin{center}
\caption{Lower and upper bounds on $M_1$ for different mass ratios of the RH neutrinos ($i=1,2$).} \label{t8}
 \begin{tabular}{|c|c|c|c|} 
\hline 
${\rm Hierarchies~\rightarrow }$&$M_{i+1}/M_i=10^2$&$M_{i+1}/M_i=10^3$ &$M_{i+1}/M_i=10^4$\\
\hline
${\rm Upper~bound~(GeV)}$&$3.64\times 10^{11}$&$3.66 \times 10^{11}$ &$3.67 \times 10^{11}$ \\
\hline
${\rm Lower~bound~~(GeV)}$&$3.55 \times 10^{11}$&$3.57 \times 10^{11}$&$3.58 \times 10^{11}$\\
\hline
\end{tabular} 
\end{center} 
\end{table}
\noindent
{\bf Direct effect of $N_{2}$:}\\  
\begin{figure}[H]
\includegraphics[scale=.45]{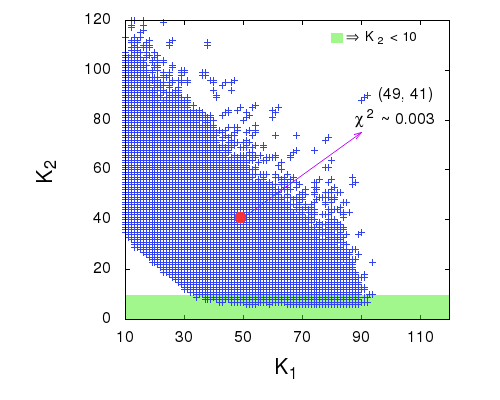} \includegraphics[scale=.45]{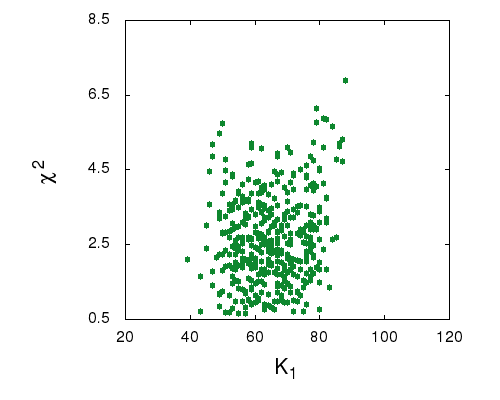}
\caption{A plot of the two washout parameters $K_1$ and $K_2$ appears in the left panel. The red dot corresponds to $\chi^2=0.003$ for which we estimate $Y_B$. The green shaded area indicates a possibility of $N_2$ leptogenesis. A plot of $\chi^2$ with $K_1$ for $K_2<10$ is given in the right panel. A normal mass ordering for the light neutrinos has been assumed.}\label{n2}
\end{figure}
\noindent
Here we consider only  $N_2$, neglecting $N_3$ for simplicity. It is argued in Ref.\cite{33} that, due to a decoherence effect\cite{5,33}, a finite lepton asymmetry generated by $N_2$ decays might remain protected against  $N_1$-washout and could survive down to the electroweak scale. Thus it itself might generate the final baryon asymmetry if a sizable amount of lepton asymmetry survives. This procedure is subject to the condition that  two washout factors $K_1$ (related to $N_1$-washout) and $K_2$ (related to $N_2$-washout) need not be of the same order. These are defined as 
\bea
K_1=\frac{\mathcal{H}_{11}}{M_1m^*},\\
K_2=\frac{\mathcal{H}_{22}}{M_2m^*},
\eea
where $m^*=1.66\sqrt{g^*}\pi v^2/M_{Pl}\approx 10^{-3}$ eV. The conditions that are needed can be stated as\cite{33}
\bea
K_1\gg 1\hspace{1mm} {\rm and}\hspace{1mm}K_2 \not\gg 1. \label{n2cond}
\eea 
Here $K_1\gg 1$ indicates that faster $N_1$ interactions break  coherence among the states produced by $N_2$, i.e. a part of the lepton asymmetry produced by $N_2$ gets protected against  $N_1$-washout. On the other hand, $K_2 \not\gg 1$ implies a mild washout of the lepton asymmetry produced by $N_2$ from $N_2$-related interactions in a way that a sizable $N_2$-generated lepton asymmetry survives during the $N_1$-leptogenesis phase. Quantitatively, our allowed parametric region (blue shaded area in the $K_2$ vs. $K_1$ plot in the left panel of Fig.\ref{n2}) prefers large values of $K_2$ in excess of 10 except at the bottom (green band). Thus the $K_2 \not\gg 1$ condition is strongly violated in most of the region. On the other hand, the few allowed points with $K_2<10$, displayed in a $\chi^2$ vs. $K_1$ plot in the right panel of Fig.\ref{n2}, correspond to values of $\chi^2$ above 0.5 far in excess of $\chi^2=0.003$ for which we obtain $Y_B$ in the observed range. Therefore, for our calculation, any direct effect of $N_2$ does not appear to be relevant.
\section{Summary and discussion }\label{s7} 
Some of us has recently proposed \cite{14} a complex-extended scaling model of the light neutrino Majorana mass matrix $M_\nu$, generated by a type-1 seesaw induced by heavy RH neutrinos. Unlike the Simple Real Scaling model advanced earlier\cite{18,19}, this new model can accommodate a nonzero $\theta_{13}$ and has a sizable region of parameter space allowed by all current and relevant experimental data\cite{27}. The atmospheric mixing angle $\theta_{23}$ is given by $\tan^{-1}(1/k)$, $k$ being a real positive scaling factor which can be either greater or less than unity. Most interesting are the predictions of the model in regard to CP violation: maximal ($\cos\delta=0$) for the Dirac type and absent ($\alpha$, $\beta$ = 0 or $\pi$) for the Majorana type. Since CP violation is crucially related to baryogenesis, we have been motivated in this paper to investigate the latter quantitatively in the model under consideration.\\

We first performed a general calculation of the CP asymmetries $\varepsilon_i^\alpha$ in the decays $N_i\rightarrow \slashed{L}_\alpha \phi,\slashed{L}^C_\alpha \phi^\dagger$ in terms of the parameters of the model. This led to a vanishing value of $\varepsilon_i^e$ with a generally nonvanishing $\varepsilon_i^\mu =-\varepsilon_i^\tau$. A common source of the origin of a nonzero $\theta_{13}$ and these CP asymmetries was found in the imaginary part of  $m_D$. We then evolved $Y_2=Y_e+Y_\mu$ and $Y_\tau$, respectively equal to $(n^{(2)}_L-n^{(2)}_{\bar{L}})/s$ and $(n^\tau_L-n^\tau_{\bar{L}})/s$,  from a high temperature (depending on the mass regime in which $M_1$ lies) down to that of the electroweak phase transition. In doing so we have had to consider the Boltzmann equations for $Y_{N_i}$ and $Y_\lambda$, respectively equal to $n_{N_i}/s$ and $(n^\lambda_L-n^\lambda_{\bar{L}})/s$, $\lambda$ being an active lepton flavor index which can sometimes be a combination of $e$, $\mu$, $\tau$.  We then utilized the different linear relations between $Y_\lambda$ and $Y_{\Delta_\lambda}$, with $\Delta_\lambda=\frac{1}{3}B-L_\lambda$, for the three different specified regimes of $M_1$ to arrive at the baryon asymmetry of the universe for each regime. The latter values have been evaluated numerically and their implications discussed.\\

 In a nutshell, realistic baryogenesis has been found to be possible in this model for values close to best fit values of the input neutrino oscillation observables only in the $10^9{\rm~GeV}<M_1<10^{12}$ GeV regime and for a normal mass ordering of the light neutrinos. This analysis  excludes (from a baryogenesis standpoint) the regimes  $M_1<10^9$ GeV and $M_1>10^{12}$ GeV and disfavors an inverted mass ordering of the light neutrinos. However, the latter is still allowed for values of the input parameters away from their best-fit numbers but within a $3\sigma$ range. As neutrino oscillation data improve, the conclusions from our analysis will be sharpened. \\

\section*{Acknowledgement}
The work of RS and AG is supported by the Department of Atomic Energy (DAE), Government of India. MC acknowledges support from Saha Institute of Nuclear Physics where a part of this work was done. The work of PR has been supported by the Indian National Science Academy.

{}

\end{document}